\documentclass[prx,aps,showpacs,twocolumn,superscriptaddress,longbibliography]{revtex4-2}
\usepackage{graphicx}
\usepackage{mathrsfs}
\usepackage{amssymb}
\usepackage{amsmath}
\usepackage{physics}
\usepackage{bbding}
\usepackage{xcolor}
\usepackage[colorlinks=true,linkcolor=blue,anchorcolor=red,citecolor=blue,urlcolor=blue]{hyperref}
\DeclareMathAlphabet\mathbfcal{OMS}{cmsy}{b}{n}

\begin{document}

\title{Probing quantum geometric nonlinear magnetization via second-harmonic magneto-optical Kerr effect}

\author{Xuan Qian}
\email{These authors contributed equally to this work.}
\affiliation{State Key Laboratory of Semiconductor Physics and Chip Technologies, Institute of Semiconductors, Chinese Academy of Sciences, Beijing 100083, China}
\affiliation{College of Materials Science and Optic-Electronic Technology, University of Chinese Academy of Sciences, Beijing 100049, China}
\author{Xiao-Bin Qiang}
\email{These authors contributed equally to this work.}
\affiliation{State Key Laboratory of Quantum Functional Materials, Department of Physics, and Guangdong Basic Research Center of Excellence for Quantum Science, Southern University of Science and Technology (SUSTech), Shenzhen 518055, China}
\author{Wenkai Zhu}
\email{These authors contributed equally to this work.}
\affiliation{State Key Laboratory of Semiconductor Physics and Chip Technologies, Institute of Semiconductors, Chinese Academy of Sciences, Beijing 100083, China}
\affiliation{College of Materials Science and Optic-Electronic Technology, University of Chinese Academy of Sciences, Beijing 100049, China}
\author{Yuqing Huang}
\affiliation{State Key Laboratory of Semiconductor Physics and Chip Technologies, Institute of Semiconductors, Chinese Academy of Sciences, Beijing 100083, China}
\affiliation{College of Materials Science and Optic-Electronic Technology, University of Chinese Academy of Sciences, Beijing 100049, China}
\author{Yiyuan Chen}
\affiliation{State Key Laboratory of Quantum Functional Materials, Department of Physics, and Guangdong Basic Research Center of Excellence for Quantum Science, Southern University of Science and Technology (SUSTech), Shenzhen 518055, China}
\affiliation{Quantum Science Center of Guangdong-Hong Kong-Macao Greater Bay Area (Guangdong), Shenzhen 518045, China}

\author{Hai-Zhou Lu}
\email{luhz@sustech.edu.cn}
\affiliation{State Key Laboratory of Quantum Functional Materials, Department of Physics, and Guangdong Basic Research Center of Excellence for Quantum Science, Southern University of Science and Technology (SUSTech), Shenzhen 518055, China}
\affiliation{Quantum Science Center of Guangdong-Hong Kong-Macao Greater Bay Area (Guangdong), Shenzhen 518045, China}
\author{Yang Ji}
\email{jiyang2024@zju.edu.cn}
\affiliation{School of Physics, Zhejiang University, Hangzhou, China}
\author{Kaiyou Wang}
\email{kywang@semi.ac.cn}
\affiliation{State Key Laboratory of Semiconductor Physics and Chip Technologies, Institute of Semiconductors, Chinese Academy of Sciences, Beijing 100083, China}
\affiliation{College of Materials Science and Optic-Electronic Technology, University of Chinese Academy of Sciences, Beijing 100049, China}

\date{\today}

\begin{abstract}
Quantum geometry provides an intrinsic framework for characterizing the geometric structure of quantum states. It highlights its relevance to various aspects of fundamental physics. However, its direct implications for magnetic phenomena remain largely unexplored. Here, we report the observation of electric-field-induced nonlinear magnetization in the nonmagnetic semimetal WTe$_2$ by using a second-harmonic magneto-optical Kerr effect (SMOKE) spectroscopy. We observe a robust nonlinear SMOKE signal that scales quadratically with current and persists up to 200 K. Theoretical modeling and scaling analysis indicate that this nonlinear magnetization is dominated by the orbital contribution and is intrinsically linked to the quantum Christoffel symbol. Just as the Christoffel symbol is a fundamental quantity encoding spacetime geometry in Einstein's general relativity, our work establishes a direct link between quantum geometry and nonlinear magnetization, and provides a geometric perspective for designing future orbitronic devices.
\end{abstract}
\maketitle

\textit{\textcolor{blue}{Introduction.}--}Quantum geometry ($\mathbf{g}-i\boldsymbol{\Omega}/2$) characterizes the distance between quantum states in parameter space~\cite{provost1980,AA90prl,resta2011}. Its imaginary part corresponds to the Berry curvature $\boldsymbol{\Omega}$, which quantifies the phase distance, while its real part defines the quantum metric $\mathbf{g}$, quantifying the amplitude distance. Recent studies have highlighted its pivotal role in a range of phenomena~\cite{torma2023,Lu24nsr,YanBH25rpp}, including flat-band superconductors~\cite{peotta2015,julku2016,lianglong2017,torma2018,huhtinen2022}, fractional Chern insulators~\cite{parameswaran2012,roy2014,jackson2015}, and nonlinear transport~\cite{GaoY14prl,FuL15prl,Lu18prl,XuSY19nature,Lu19nc,GaoY21prl,YangSY21prl,XuSY23science,GaoWB23nature,Lu25as}. In the area of magnetic responses~\cite{Edestein90ssc,Kato04prl,ZhangSC05prl,Inoue08prb,Leonid09np,Ferguson11nn,Dimi07prl,Sinowa14prl,Manchon19rmp,Yuriy21epl,Felix24rmp,Jo24npj,WangP24aem,Lu26apl}, the Berry curvature provides a well-established description of linear magnetization ($\mathbf{M}\propto\mathbf{E}$, where $\mathbf{M}$ is the magnetization and $\mathbf{E}$ is the electric field)~\cite{Niu05prl,Resta05prl,Resta06prb,Niu07prl,Xiao10rmp,Yoda15sr,Moore16prl,Pesin17prb,Lee19prl,Gorden19science,Murakami20prb,Johansson24jpcm}. Nevertheless, the influence of the quantum metric $\mathbf{g}$ on magnetic responses, especially in the nonlinear regime ($\mathbf{M}\propto\mathbf{E}\mathbf{E}$), remains largely unexplored. This gap stems from the difficulty of detecting weak magnetic signals and the lack of a systematic theory for higher-order effects.

\begin{figure}[htbp]
\centering 
\includegraphics[width=0.49\textwidth]{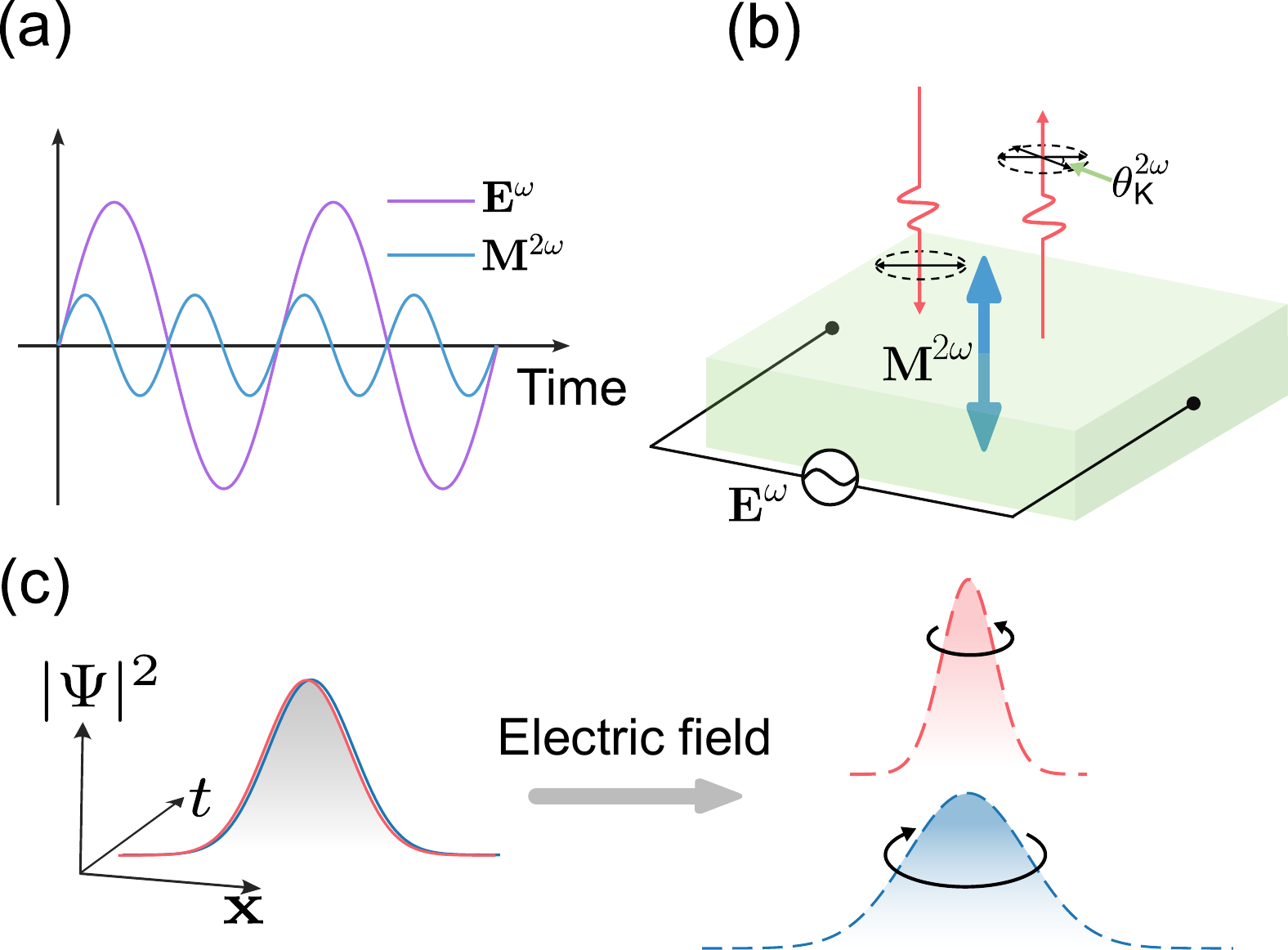}
\caption{SMOKE detection and quantum geometric nonlinear magnetization. (a) Demonstration of second-harmonic nonlinear magnetization $\mathbf{M}^{2\omega}$ at frequency $2\omega$ induced by an alternating electric field $\mathbf{E}^{\omega}$ at frequency $\omega$, i.e., ($\mathbf{M}^{2\omega}\propto\mathbf{E}^\omega\mathbf{E}^\omega$). (b) Schematic of SMOKE detection to the nonlinear magnetization, the red downward (upward) arrow represents incident (reflected) light, the green arrow represents induced magnetization with frequency $2\omega$, and purple arrow represents applied alternating electric field with frequency $\omega$. (c) Schematic of nonlinear magnetization origin. Orbital magnetization arises from the self-rotation of Bloch wavepacket (circular arrows)~\cite{Xiao10rmp,GaoY14prl}. An applied electric field modifies the wave packets (solid to dashed curves) and induces a nonequilibrium distribution. This prevents the mutual cancellation of time-reversal counterparts (red and blue), and leads to a quadratic electric-field dependence.} 
\label{Fig: demo} 
\end{figure}

In this Letter, we report the observation of electric-field-induced nonlinear magnetization in the nonmagnetic semimetal WTe$_2$. This was achieved by developing a high-sensitivity SMOKE spectroscopy that employs lock-in detection to isolate the second-harmonic Kerr signals from background noise [Figs.~\ref{Fig: demo}(a) and (b)].
Therefore, this technique extends conventional magneto-optical Kerr effect (MOKE)~\cite{Cowburn03jpd,Kato2004,Thomas09np,Stamm17prl,Choi23nature,Lyalin23prl} into the nonlinear regime. We observed a robust SMOKE signal that scales quadratically with the applied electric field up to 200 K. Symmetry analysis and theoretical calculations further reveal that this nonlinear magnetization is dominated by orbital contribution, which can be heuristically understood through the field-induced deformation of Bloch wavepackets [Fig.~\ref{Fig: demo}(c)].  Strikingly, this deformation is fundamentally governed by a geometric quantity, the quantum Christoffel symbol, which is directly related to the quantum metric $\mathbf{g}$ [Eq.~\eqref{Eq: Gamma}]. The additional scaling analysis shows that the experimental data are in good agreement with theoretical predictions across a broad range of temperatures.

\begin{figure}[htbp]
\centering 
\includegraphics[width=0.49\textwidth]{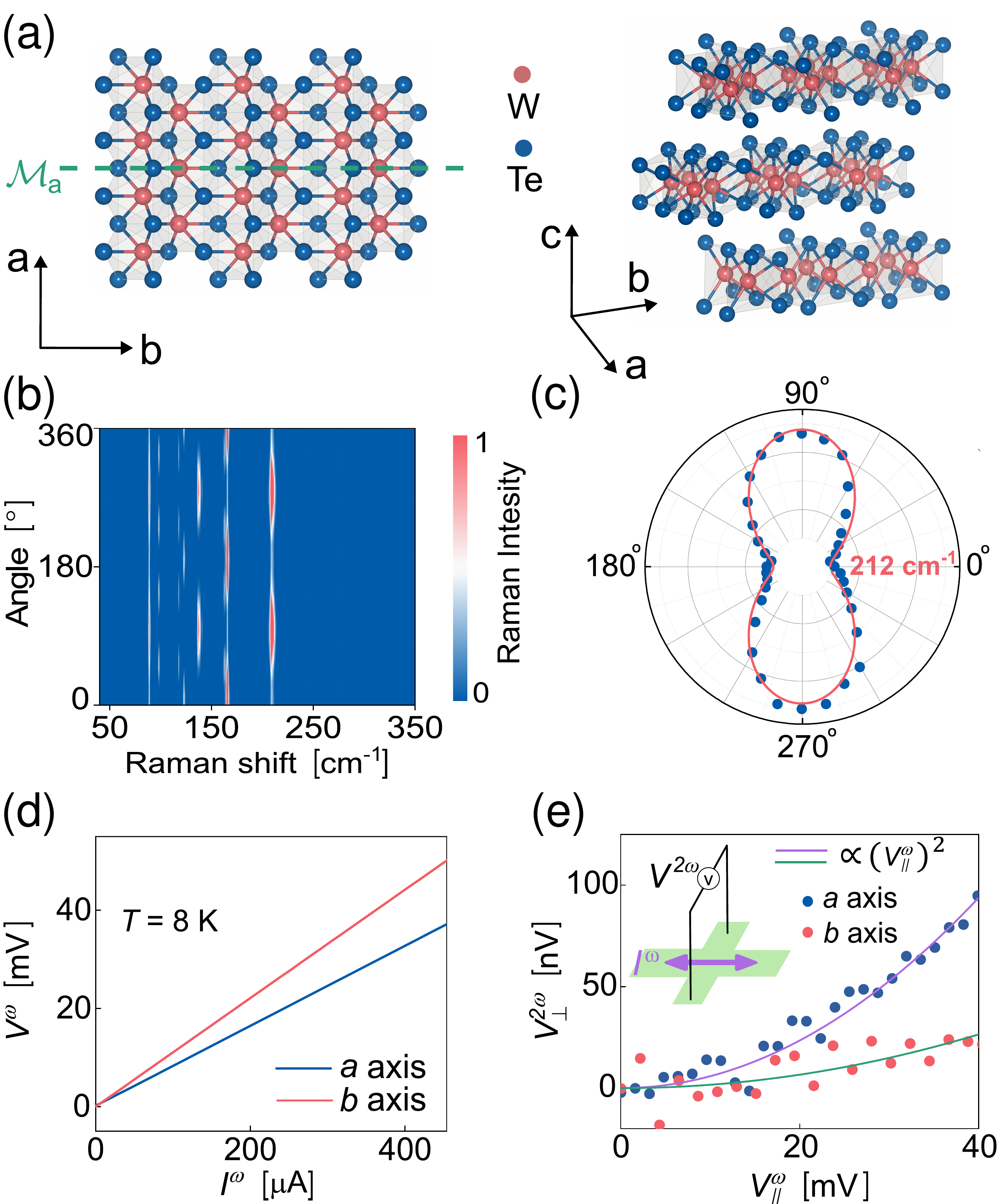}
\caption{Crystal structure and electrical transport properties of WTe$_2$. (a) The crystal structure of WTe$_2$. The ideal crystal structure of Td-phase WTe$_2$ only preserves a mirror symmetry $\mathcal{M}_a$ (with the mirror plane indicated by the dashed line). (b) Angle-dependent polarized Raman spectral intensity of a WTe$_2$ flake used in the device. The measurements were obtained by rotating the laser polarization with respect to the \textit{a} axes. (c) Representative angle-dependent intensities of the Raman peak at 212 cm$^{-1}$. (d) First-harmonic longitudinal response $V^\omega$ with the alternating current applied along the \textit{a} and \textit{b} axes. (e) Second-harmonic transverse response $V_{\perp}^{2\omega}$ with the applied alternating current along the \textit{a} and \textit{b} axes (the horizontal coordinate has been converted to the voltage). The dots represent the experimental data, and the solid lines represent the fitted quadratic curves. The inset in (e) shows the measurement setup. Data in (d) and (e) were collected at $T = 8$ K.}
\label{Fig: lat}
\end{figure}

\begin{figure*}[htbp]
\centering
\includegraphics[width=0.75\textwidth]{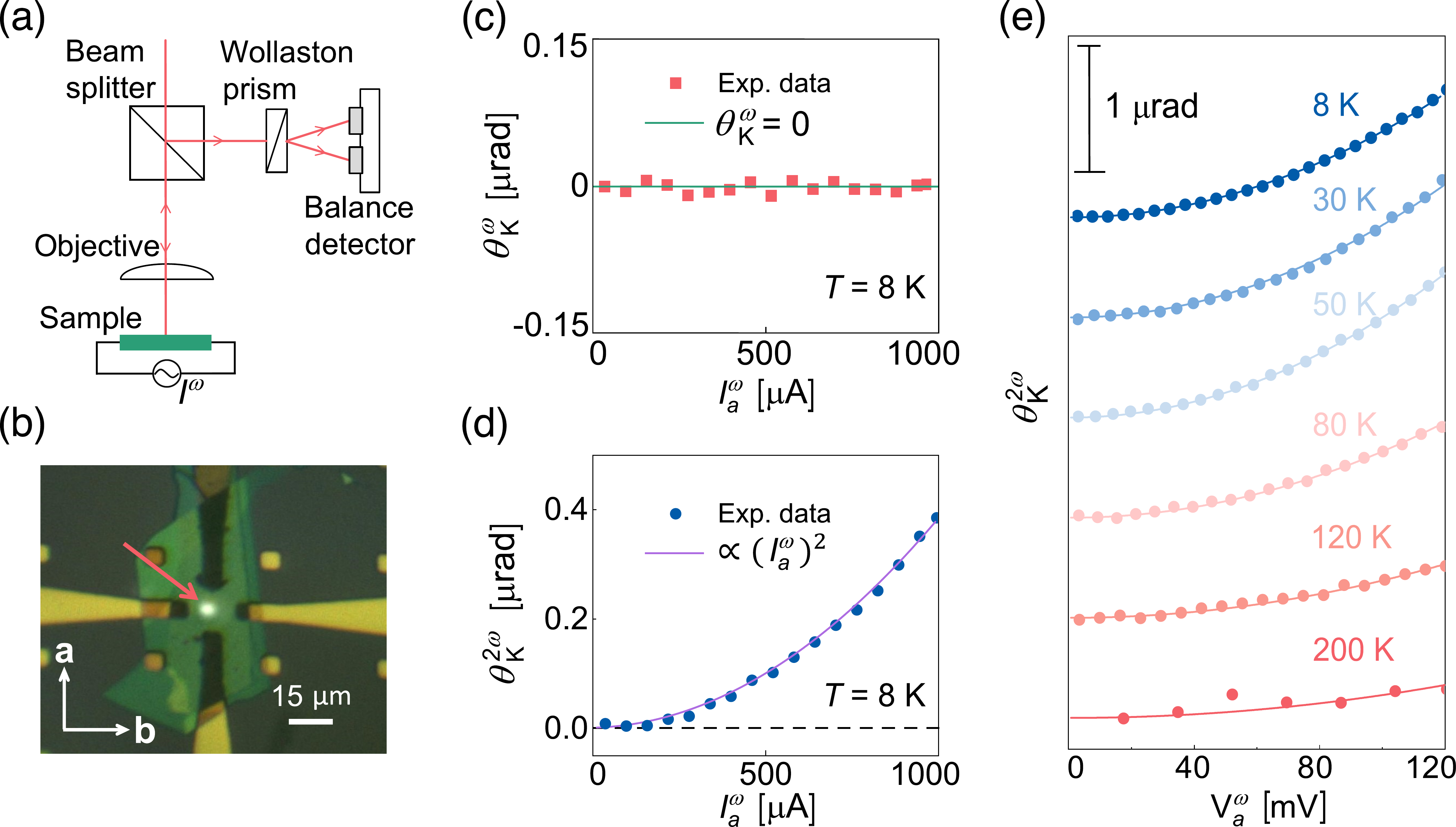} 
\caption{Observation of SMOKE and temperature-dependence. (a) Optical path for polar Kerr measurements. (b) Optical image of the WTe$_2$ device. The red arrow indicates the light spot. (c) The MOKE signal $\theta_{\text{K}}^{\omega}$ as a function of current $I_a^{\omega}$, and there is no signal within the measurement accuracy. (d) The SMOKE signal $\theta_{\text{K}}^{2\omega}$ as a function of current $I_{a}^{\omega}$, the solid line represents a quadratic fit to the data. (e) Temperature dependence of the SMOKE from 8 K to 200 K.}
\label{Fig: exp}
\end{figure*}

\textit{\textcolor{blue}{Demonstration of mirror symmetry breaking.}--}According to the symmetry analysis presented in Supplemental Material SI~\cite{Supp}, an out-of-plane nonlinear magnetization $M_c$ induced by an in-plane electric field $E_a$ requires breaking both the two-fold rotational symmetries ($\mathcal{C}_a^2$ and $\mathcal{C}_b^2$) and the mirror symmetries ($\mathcal{M}_a$ and $\mathcal{M}_b$), where $(a,b,c)$ represent crystallographic directions. This requirement suggests that multilayer Td-phase WTe$_2$, a nonmagnetic semimetal, could serve as a promising candidate. As shown in Fig.~\ref{Fig: lat}(a), ideal multilayer WTe$_2$ preserves only a single mirror symmetry $\mathcal{M}_a$. However, in realistic devices, this residual symmetry is typically weakly broken due to interface and strain effects~\cite{Puri24NL,LiaoZM24prb2,LI25MT,Zhang25nc}. To determine whether $\mathcal{M}_a$ is similarly broken in our devices, we utilize the nonlinear Hall effect as a probe. Since $\mathcal{M}_a$ also forbids the nonlinear Hall response when the current is applied along the $b$ axis~\cite{FuL15prl,Lu18prl,XuSY19nature}, the observation of such a signal can provide direct evidence of $\mathcal{M}_a$ breaking.

For the nonlinear Hall measurements, we utilized a device incorporating a $\sim$7 nm thick Td-phase WTe$_2$ flake (Supplemental Material SII~\cite{Supp}). The crystallographic \textit{a} axis was identified by combining optical microscopy of the flake edges with high-precision polarized Raman spectroscopy [Figs.~\ref{Fig: lat}(b) and (c); Supplemental Material SIII~\cite{Supp}]. As shown in Fig.~\ref{Fig: lat}(d), the linear I-V curves confirm high-quality Ohmic contacts. We then measured the nonlinear Hall responses along both the \textit{a} and \textit{b} axes using the lock-in technique (Supplemental Material SIV~\cite{Supp}). Clear signals of the nonlinear Hall effect were observed in both directions [Fig.~\ref{Fig: lat}(e)]. Notably, although the \textit{b}-axis response is weaker than the \textit{a}-axis, its presence serves as robust evidence for mirror symmetry ($\mathcal{M}_a$) breaking. This establishes the Td-phase WTe$_2$ as an ideal platform to search for the nonlinear magnetization.

\textit{\textcolor{blue}{Observation of nonlinear magnetization via SMOKE.}--}Using the same device as in the nonlinear Hall transport measurements, we carried out the optical measurements. Figure~\ref{Fig: exp}(a) depicts the polar Kerr setup for detecting out-of-plane magnetization. A linearly polarized laser beam was normally incident and focused onto the sample with applied a alternating current ($\omega$=1.177 kHz). The Kerr rotation of the reflected light was collected by a balanced detection system. A lock-in amplifier then demodulated this signal at the first-harmonic frequency ($\omega$=1.177 kHz) to extract MOKE response and at the second-harmonic frequency ($2\omega$=2.354 kHz) to isolate the SMOKE signal (see Supplemental Material SV~\cite{Supp} for more details). To ensure consistency when switching the current direction between crystallographic axes, the light spot was precisely positioned at the central intersection of the \textit{a} and \textit{b} axes [Fig.~\ref{Fig: exp}(b)]. Consequently, by combining a stabilized tunable laser system with real-time intensity monitoring and phase-sensitive lock-in detection, we achieved an ultra-high Kerr rotation resolution of down to 10 nrad. 

As shown in Fig.~\ref{Fig: exp}(c) and Supplemental Fig.~S6(a), MOKE measurements at 8 K show no detectable signal with current applied along either the \textit{a} or \textit{b} axis. In contrast, a pronounced SMOKE signal $\theta_{\text{K}}^{2\omega}$ emerges and exhibits a clear quadratic dependence on the applied current $I^{\omega}$ [Fig.~\ref{Fig: exp}(d) and Supplemental Fig.~S6(b)], confirming the second-order nonlinear nature of the effect. Additionally, to examine the robustness of this effect, we measured the temperature dependence of the SMOKE signal. As shown in Fig.~\ref{Fig: exp}(e), $\theta_{\text{K}}^{2\omega}$ maintains a clear and robust presence across a broad temperature range from 8 to 200~K. This observation is particularly confirms the high thermal stability of the nonlinear magnetization.

To rule out the alternative origins of the SMOKE signal beyond the nonlinear magnetization, we performed a series of control experiments. First, we confirmed the reproducibility across multiple WTe$_2$ devices from different batches. All devices exhibited a clear SMOKE signal with no detectable MOKE response, and in most cases, the SMOKE signal persisted up to 200~K (Supplemental Fig.~S7~\cite{Supp}). Second, to exclude current-induced heating effect, a graphene control device of similar thickness was measured under the same current range, but no SMOKE signal was observed (see Supplemental Material SVI~\cite{Supp}). Particularly, graphene possesses one of the largest third-order nonlinear susceptibilities among all known two-dimensional materials~\cite{Xie24small}, making the absence of a SMOKE signal decisive in ruling out a second-order electro‑optic effect origin. Finally, power‑dependent SMOKE measurements showed no variation with laser intensity, eliminating laser‑induced heating effect (Supplemental Material SVII~\cite{Supp}). Together, these control experiments confirm that the observed SMOKE signal in WTe$_2$ originates from an electric‑field‑induced nonlinear magnetization.

\begin{figure*}[htbp]
\centering 
\includegraphics[width=0.7\textwidth]{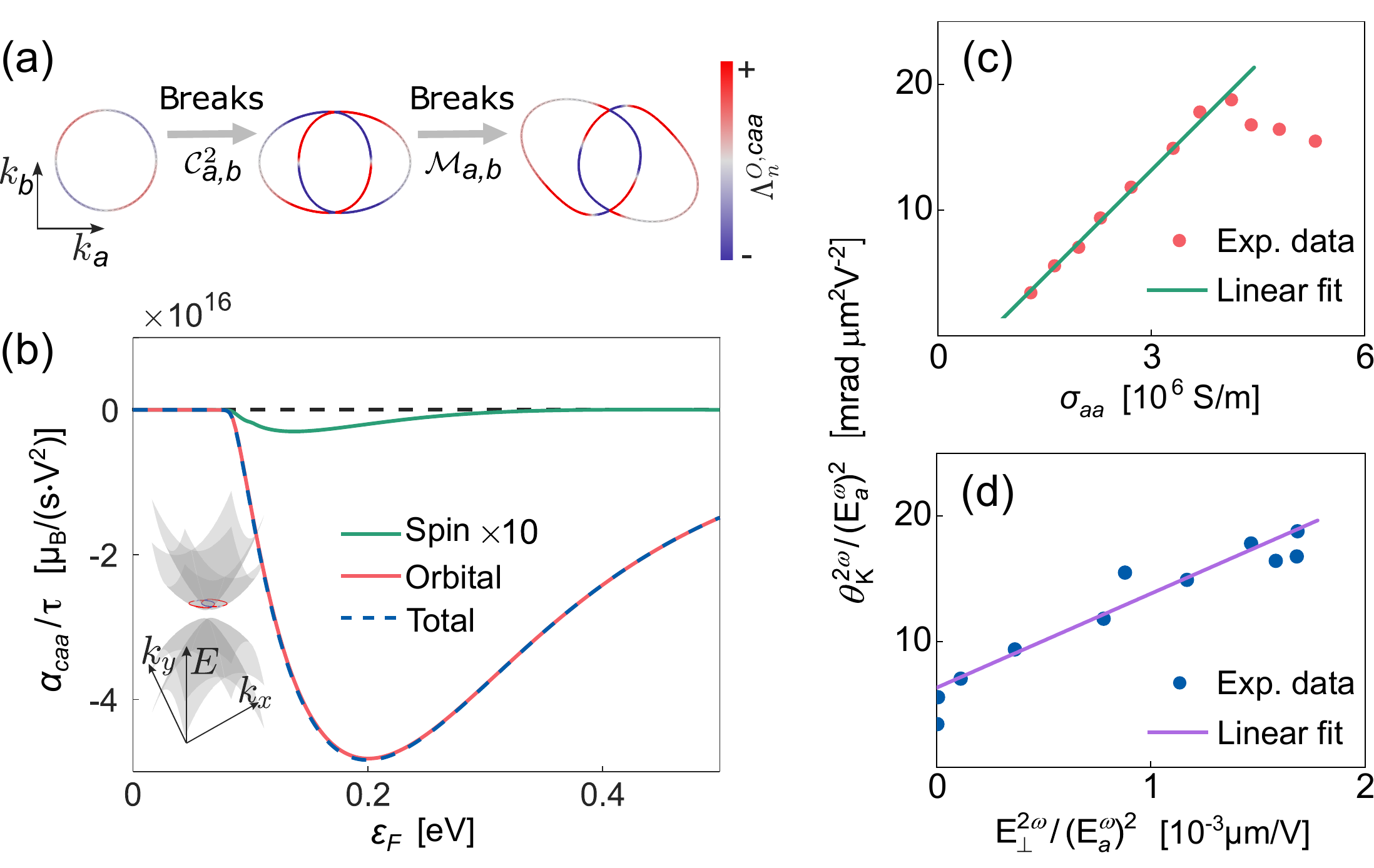}
\caption{Theoretical results and scaling analysis. (a) The evolution of Fermi surfaces at Fermi level $\varepsilon_F=0.25$ eV by breaking two-fold rotational symmetries ($\mathcal{C}_{a,b}^2$) and mirror symmetries ($\mathcal{M}_{a,b}$) successively, the colorbar depicts the strength of $\Lambda_n^{\text{O},caa}$. (b) Calculated nonlinear magnetization coefficient $\alpha_{caa}$ as a function of the Fermi energy $\varepsilon_F$. The green, red, and blue represent spin ($\times$10 for visibility), orbital, and total contributions, respectively. The inset depicts the energy dispersion. The model parameters are $v=1$ eV$\cdot$nm, $t/v=0.3$, $m=0.1$ eV, and $b=1$ eV$\cdot$nm$^2$~\cite{Lu18prl}. (c) Scaling behavior between the observed SMOKE signal and linear longitudinal conductivity for different temperatures from $T=8$ K to $T=200$ K.  (d) Scaling behavior between the observed SMOKE signal and nonlinear Hall signal. The two data points lie on the vertical axis because the nonlinear Hall signal becomes undetectable at $T=150$ K and 200 K.}
\label{Fig: theo}
\end{figure*}

\textit{\textcolor{blue}{Quantum geometric origin of nonlinear magnetization.}--}The nonlinear magnetization induced by a low-frequency ($\sim$kHz) alternating electric field can be generally expressed as
\begin{equation}\label{Eq: M_EE}
\mathbf{M}^{2\omega}(t)=\boldsymbol{\alpha}\mathbf{E}^\omega(t)\mathbf{E}^\omega(t),
\end{equation}
where $\mathbf{M}^{2\omega}(t) = \text{Re}\{\mathbf{M} e^{i2\omega t}\}$ is the magnetization oscillating at frequency $2\omega$, $\mathbf{E}^\omega(t) = \text{Re}\{\mathbf{E} e^{i\omega t}\}$ is the electric field applied at frequency $\omega$, and $\boldsymbol{\alpha} = \boldsymbol{\alpha}^{\textbf{S}} + \boldsymbol{\alpha}^{\textbf{O}}$ is the nonlinear magnetization coefficient (a rank-3 tensor), with S and O denoting spin and orbital contributions.

By employing Boltzmann kinetics~\cite{Haug08book} in combination with wavepacket dynamics~\cite{Xiao10rmp,GaoY14prl}, the nonlinear magnetization coefficient $\alpha_{ijk}$ for a time-reversal-symmetric (nonmagnetic) system can be written as (check Supplemental Material SVIII~\cite{Supp} for details)
\begin{equation}\label{Eq: alpha}
\alpha_{ijk}=\tau e^2\int[d\mathbf{k}] \Lambda_n^{ijk} f_0',
\end{equation}
where $(i,j,k)$ denote the crystallographic directions $(a,b,c)$, $\tau$ is the relaxation time of conduction electrons, $-e$ is electron charge, $[d\mathbf{k}]$ represents $ \sum_n d^d\mathbf{k}/(2\pi)^d$ with dimension $d$ and band index $n$, and $f_0'$ denotes the derivative of the Fermi–Dirac distribution with respect to the band energy. The full geometric information of the system is encoded in the quantity $\Lambda_n^{ijk}$, it is defined by $\Lambda_n^{ijk}=(F_n^{\text{S},ij}+F_n^{\text{O},ij})v_n^k$, where $\mathbf{F}_n^{\text{S/O}}$ is the spin/orbital polarizability tensor for the $n$th band~\cite{GaoY14prl,YangSY24prl}, and $\mathbf{v}_n$ is the intraband velocity. Explicitly, the expressions of $F_n^{\text{S},ij}$ and $F_n^{\text{O},ij}$ are given by
\begin{equation}\label{Eq: Fn}
\left\{
\begin{aligned}
F_n^{\text{S}, ij}&=2g\mu_B\text{Im} \sum_{m \neq n} \frac{\langle n|s_i|m \rangle \langle m|\partial_j\mathcal{H}|n \rangle}{(\varepsilon_n -\varepsilon_m)^2},\\
F_n^{\text{O}, ij}&=-2\text{Im} \sum_{m \neq n} \frac{ m_{nm}^{\text{O},i} \langle m|\partial_j \mathcal{H}|n\rangle}{(\varepsilon_n - \varepsilon_m)^2}
-\frac{e}{2\hbar}\epsilon_{ikl}\Gamma_n^{lkj},
\end{aligned}\right.
\end{equation}
where $g$ is the Land\'{e} $g$-factor for spin, $\mu_B$ is the Bohr Magneton, $\hbar$ is the reduced Planck constant, $\mathbf{s}=\boldsymbol{\sigma}/2$ is the dimensionless spin operator, $\ket{n}$ is the Bloch state, $\mathcal{H}$ is the Bolch Hamiltonian, $\varepsilon_n$ is the band energy, $\mathbf{m}_{mn}^{\text{O}}=\sum_{l\neq n}(\mathbf{v}_{m l}+\delta_{lm }\mathbf{v}_n)\times \mathbfcal{A}_{ln}/2$ is the interband orbital magnetic moment, $\mathbf{v}_{mn}=\langle m| \partial_\mathbf{k} \mathcal{H} | n \rangle/\hbar$ is the interband velocity, $\mathbfcal{A}_{mn}=\bra{m}i\partial_\mathbf{k}\ket{n}$ is the interband Berry connection, $\epsilon_{ijk}$ is the Levi-Civita symbol, and $\partial_i$ is the short of $\partial/(\partial k_i)$. Notably, the second term of $F_n^{\text{O},ij}$ arises from the quantum Christoffel symbol
\begin{eqnarray}\label{Eq: Gamma}
\Gamma_n^{lkj} = \frac{1}{2}(\partial_j g_n^{kl} + \partial_k g_n^{lj} - \partial_l g_n^{kj}),  
\end{eqnarray}
where $g_n^{ij}=\text{Re} \sum_{m\neq n}\mathcal{A}_{nm}^i \mathcal{A}_{mn}^j$ is the quantum metric tensor~\cite{provost1980,AA90prl,resta2011}. In Einstein's general relativity, the Christoffel symbol originally characterizes curved spacetime through the geodesic equation~\cite{KPThorne73book}. Recently, its quantum version (quantum Christoffel symbol) has been introduced to describe momentum-space gravity~\cite{Smith22prr,Mehraeen25prl,FuL26prb}.

\textit{\textcolor{blue}{Effective model calculations and scaling analysis.}--}To obtain an intuitive understand of nonlinear magnetization, we establish an effective model that inherits the same symmetry constraints as WTe$_2$. The electronic structure of WTe$_2$ is a type-II semimetal with strong spin-orbit coupling~\cite{DaiX15nature,WanXG16nc,XuSY18np}. Therefore, the modified Dirac model~\cite{Lu18prl,Lu23prb} can give the ingredient
\begin{equation}\label{Eq: model}
\mathcal{H} = v (k_a \alpha_a + k_b \alpha_b) + (m-bk^2) \beta+\mathcal{H}',
\end{equation} 
where $v,m,b$ are model parameters, $\alpha_i=\sigma_a \otimes \sigma_i$, $\beta=\sigma_c \otimes \sigma_0$ are Dirac matrices defined within Pauli matrices $(\sigma_a,\sigma_b,\sigma_c)$. To satisfy the symmetry constraints of WTe$_2$, a symmetry-breaking term $\mathcal{H}' = t k_a \sigma_c \otimes (\sigma_a + \sigma_b + \sigma_c)$ is introduced to break the two-fold rotational symmetries $\mathcal{C}_{a,b}^2$ and the mirror symmetries $\mathcal{M}_{a,b}$ but preserve the time-reversal symmetry. Figure~\ref{Fig: theo}(a) demonstrates the evolution of Fermi surface by breaking $\mathcal{C}_{a,b}^2$ and $\mathcal{M}_{a,b}$ symmetries successively. Upon symmetry breaking, the Fermi surface exhibits an non-symmetric distribution of $\Lambda_n^{\text{O},caa}$ in $\mathbf{k}$-space, which coincides the symmetry requirements for a nonzero $\alpha_{caa}$.

Using Eqs.~\eqref{Eq: alpha} and \eqref{Eq: Fn}, we calculate the nonlinear magnetization coefficient $\alpha_{caa}$ for the effective model of WTe$_2$ [Eq.~\eqref{Eq: model}]. Figure~\ref{Fig: theo}(b) shows the calculated $\alpha_{caa}$, including spin, orbital, and total contributions, as a function of the Fermi energy $\varepsilon_F$. Notably, across the entire range of Fermi energy, the orbital contribution remains dominant. To be specific, the orbital contribution is approximately two orders of magnitude larger than the spin contribution. In our experiment, the applied electric field is around $5\times10^3$ V/m. Then, the induced nonlinear spin magnetization is $5.42\times 10^{-9}$ $\mu_B/$nm$^2$, and the orbital magnetization is $1.32\times 10^{-6}$ $\mu_B/$nm$^2$ for $\varepsilon_F=0.2$ eV, $t/v=0.3$, and $\tau=1.1$ ps (the estimation of scattering time can be found in Supplemental Material SIX~\cite{Supp}). Therefore, this magnitude of magnetization is sufficient to be detected by our high-precision SMOKE spectroscopy.

Scaling analysis is a powerful tool for distinguishing experimental mechanisms, as demonstrated in studies of the anomalous and nonlinear Hall effects~\cite{Nagaosa10rmp,Lu19nc,GongZH24arXiv,GongZH25arXiv}. This method assumes that, aside from Fermi-surface broadening, the temperature dependence in metals arises predominantly from the relaxation time $\tau$. Since the longitudinal conductivity $\sigma_{aa}\propto\tau$, other response coefficients thus can be expressed as a polynomial in $\sigma_{aa}$. Therefore, the specific power-law dependence on $\sigma_{aa}$ as the temperature varies allows for the determination of the dominant mechanism.

Eq.~\eqref{Eq: alpha} predicts that the quantum geometric nonlinear magnetization is proportional to $\tau$, which implies that the coefficient $\alpha_{caa}$ should scale linearly with the longitudinal conductivity $\sigma_{aa}$. This is confirmed by our experimental data in Fig.~\ref{Fig: theo}(c) over a broad temperature range. The slight deviation observed below 20 K in Fig.~\ref{Fig: theo}(c) is attributed to the proximity of the Fermi level to the band edge in WTe$_2$~\cite{DaiX15nature,WanXG16nc,XuSY18np}. Because the quantum-geometric contribution is strongly concentrated near the band edge [Fig.~\ref{Fig: theo}(b)], Fermi-surface broadening becomes non-negligible at these temperatures. This interpretation is further supported by the linear scaling between the SMOKE signal and the nonlinear Hall signal [Fig.~\ref{Fig: theo}(d)], as both effects share the same relaxation-time dependence and a similar band-geometry origin~\cite{Lu18prl,XuSY19nature}. Taken together, the scaling analyses in Figs.~\ref{Fig: theo}(c) and (d) demonstrate strong mutual consistency between our experiments and theoretical analysis.

\textit{\textcolor{blue}{Discussion and conclusion.}--}Although the nonlinear Hall effect and the nonlinear magnetization share some similarities, they are fundamentally distinct nonlinear responses. The nonlinear Hall effect is an electric-field–induced charge response governed by the Berry-curvature dipole. In contrast, the nonlinear magnetization is a electric-field–induced magnetic response related to the quantum Christoffel symbol. This distinction is also reflected on the experimental side. The nonlinear Hall effect is readily detected via standard transport techniques, while probing nonlinear magnetization requires new measurement capabilities. The SMOKE measurement established here offers a powerful optical route for detecting weak magnetic signals. Additionally, its inherent spatial resolution further provides a foundation for more refined and spatially resolved measurements in future studies.

To summarize, our development of high-sensitivity SMOKE spectroscopy has enabled the direct observation of electric-field-induced nonlinear magnetization in the nonmagnetic semimetal WTe$_2$. The observed SMOKE signal persists up to 200 K and displays a quadratic dependence on the applied current, highlighting its robust nonlinear nature. Theoretical analysis and scaling law identify the orbital contribution as the dominant mechanism to this nonlinear magnetization. Notably, this orbital contribution is intrinsically related to the quantum Christoffel symbol, thus establishes a direct link between quantum geometry and nonlinear magnetization. Our work offers a experimental paradigm and establishes a fundamental framework for characterizing and understanding nonlinear magnetization. It paves a way toward next-generation orbitronic devices and opens an avenue for exploring emergent gravity effects in the quantum materials.

\begin{acknowledgments}
The authors are indebted to Hao-Jie Lin, Erding Zhao and Xun Yang for the insightful discussions. This work was supported by the National Key R\&D Program of China (2022YFA1405100 
and 2022YFA1403700), 
the National Natural Science Foundation of China (12427805, 12241405, 
12350402, 12525401, 
12274406, 
12174378, 
12404146, 
and 12374077), 
CAS Project for Young Scientists in Basic Research (No.~YSBR-120), 
Guangdong Basic and Applied Basic Research Foundation (2023B0303000011), Guangdong Provincial Quantum Science Strategic Initiative (GDZX2201001 and GDZX2401001), the Science, Technology and Innovation Commission of Shenzhen Municipality (ZDSYS20190902092905285), High-level Special Funds (G03050K004), the New Cornerstone Science Foundation through the XPLORER PRIZE, and Center for Computational Science and Engineering of SUSTech.
\end{acknowledgments}


\begin{thebibliography}{81}%
	\makeatletter
	\providecommand \@ifxundefined [1]{%
		\@ifx{#1\undefined}
	}%
	\providecommand \@ifnum [1]{%
		\ifnum #1\expandafter \@firstoftwo
		\else \expandafter \@secondoftwo
		\fi
	}%
	\providecommand \@ifx [1]{%
		\ifx #1\expandafter \@firstoftwo
		\else \expandafter \@secondoftwo
		\fi
	}%
	\providecommand \natexlab [1]{#1}%
	\providecommand \enquote  [1]{``#1''}%
	\providecommand \bibnamefont  [1]{#1}%
	\providecommand \bibfnamefont [1]{#1}%
	\providecommand \citenamefont [1]{#1}%
	\providecommand \href@noop [0]{\@secondoftwo}%
	\providecommand \href [0]{\begingroup \@sanitize@url \@href}%
	\providecommand \@href[1]{\@@startlink{#1}\@@href}%
	\providecommand \@@href[1]{\endgroup#1\@@endlink}%
	\providecommand \@sanitize@url [0]{\catcode `\\12\catcode `\$12\catcode
		`\&12\catcode `\#12\catcode `\^12\catcode `\_12\catcode `\%12\relax}%
	\providecommand \@@startlink[1]{}%
	\providecommand \@@endlink[0]{}%
	\providecommand \url  [0]{\begingroup\@sanitize@url \@url }%
	\providecommand \@url [1]{\endgroup\@href {#1}{\urlprefix }}%
	\providecommand \urlprefix  [0]{URL }%
	\providecommand \Eprint [0]{\href }%
	\providecommand \doibase [0]{https://doi.org/}%
	\providecommand \selectlanguage [0]{\@gobble}%
	\providecommand \bibinfo  [0]{\@secondoftwo}%
	\providecommand \bibfield  [0]{\@secondoftwo}%
	\providecommand \translation [1]{[#1]}%
	\providecommand \BibitemOpen [0]{}%
	\providecommand \bibitemStop [0]{}%
	\providecommand \bibitemNoStop [0]{.\EOS\space}%
	\providecommand \EOS [0]{\spacefactor3000\relax}%
	\providecommand \BibitemShut  [1]{\csname bibitem#1\endcsname}%
	\let\auto@bib@innerbib\@empty
	\bibitem [{\citenamefont {Provost}\ and\ \citenamefont
		{Vallee}(1980)}]{provost1980}%
	\BibitemOpen
	\bibfield  {author} {\bibinfo {author} {\bibfnamefont {J.}~\bibnamefont
			{Provost}}\ and\ \bibinfo {author} {\bibfnamefont {G.}~\bibnamefont
			{Vallee}},\ }\bibfield  {title} {\bibinfo {title} {Riemannian structure on
			manifolds of quantum states},\ }\href {https://doi.org/10.1007/BF02193559}
	{\bibfield  {journal} {\bibinfo  {journal} {Commun. Math. Phys.}\ }\textbf
		{\bibinfo {volume} {76}},\ \bibinfo {pages} {289} (\bibinfo {year}
		{1980})}\BibitemShut {NoStop}%
	\bibitem [{\citenamefont {Anandan}\ and\ \citenamefont
		{Aharonov}(1990)}]{AA90prl}%
	\BibitemOpen
	\bibfield  {author} {\bibinfo {author} {\bibfnamefont {J.}~\bibnamefont
			{Anandan}}\ and\ \bibinfo {author} {\bibfnamefont {Y.}~\bibnamefont
			{Aharonov}},\ }\bibfield  {title} {\bibinfo {title} {Geometry of quantum
			evolution},\ }\href {https://doi.org/10.1103/PhysRevLett.65.1697} {\bibfield
		{journal} {\bibinfo  {journal} {Phys. Rev. Lett.}\ }\textbf {\bibinfo
			{volume} {65}},\ \bibinfo {pages} {1697} (\bibinfo {year}
		{1990})}\BibitemShut {NoStop}%
	\bibitem [{\citenamefont {Resta}(2011)}]{resta2011}%
	\BibitemOpen
	\bibfield  {author} {\bibinfo {author} {\bibfnamefont {R.}~\bibnamefont
			{Resta}},\ }\bibfield  {title} {\bibinfo {title} {The insulating state of
			matter: a geometrical theory},\ }\href
	{https://doi.org/10.1140/epjb/e2010-10874-4} {\bibfield  {journal} {\bibinfo
			{journal} {Eur. Phys. J. B}\ }\textbf {\bibinfo {volume} {79}},\ \bibinfo
		{pages} {121} (\bibinfo {year} {2011})}\BibitemShut {NoStop}%
	\bibitem [{\citenamefont {T\"orm\"a}(2023)}]{torma2023}%
	\BibitemOpen
	\bibfield  {author} {\bibinfo {author} {\bibfnamefont {P.}~\bibnamefont
			{T\"orm\"a}},\ }\bibfield  {title} {\bibinfo {title} {{Essay: Where Can
				Quantum Geometry Lead Us?}},\ }\href
	{https://doi.org/10.1103/PhysRevLett.131.240001} {\bibfield  {journal}
		{\bibinfo  {journal} {Phys. Rev. Lett.}\ }\textbf {\bibinfo {volume} {131}},\
		\bibinfo {pages} {240001} (\bibinfo {year} {2023})}\BibitemShut {NoStop}%
	\bibitem [{\citenamefont {Liu}\ \emph {et~al.}(2024)\citenamefont {Liu},
		\citenamefont {Qiang}, \citenamefont {Lu},\ and\ \citenamefont
		{Xie}}]{Lu24nsr}%
	\BibitemOpen
	\bibfield  {author} {\bibinfo {author} {\bibfnamefont {T.}~\bibnamefont
			{Liu}}, \bibinfo {author} {\bibfnamefont {X.-B.}\ \bibnamefont {Qiang}},
		\bibinfo {author} {\bibfnamefont {H.-Z.}\ \bibnamefont {Lu}},\ and\ \bibinfo
		{author} {\bibfnamefont {X.~C.}\ \bibnamefont {Xie}},\ }\bibfield  {title}
	{\bibinfo {title} {Quantum geometry in condensed matter},\ }\href
	{https://doi.org/10.1093/nsr/nwae334} {\bibfield  {journal} {\bibinfo
			{journal} {Natl. Sci. Rev.}\ }\textbf {\bibinfo {volume} {12}},\ \bibinfo
		{pages} {nwae334} (\bibinfo {year} {2024})}\BibitemShut {NoStop}%
	\bibitem [{\citenamefont {Jiang}\ \emph {et~al.}(2025)\citenamefont {Jiang},
		\citenamefont {Holder},\ and\ \citenamefont {Yan}}]{YanBH25rpp}%
	\BibitemOpen
	\bibfield  {author} {\bibinfo {author} {\bibfnamefont {Y.}~\bibnamefont
			{Jiang}}, \bibinfo {author} {\bibfnamefont {T.}~\bibnamefont {Holder}},\ and\
		\bibinfo {author} {\bibfnamefont {B.}~\bibnamefont {Yan}},\ }\bibfield
	{title} {\bibinfo {title} {{Revealing quantum geometry in nonlinear quantum
				materials}},\ }\href {https://doi.org/10.1088/1361-6633/ade454} {\bibfield
		{journal} {\bibinfo  {journal} {Rep. Prog. Phys.}\ }\textbf {\bibinfo
			{volume} {88}},\ \bibinfo {pages} {076502} (\bibinfo {year}
		{2025})}\BibitemShut {NoStop}%
	\bibitem [{\citenamefont {Peotta}\ and\ \citenamefont
		{T{\"o}rm{\"a}}(2015)}]{peotta2015}%
	\BibitemOpen
	\bibfield  {author} {\bibinfo {author} {\bibfnamefont {S.}~\bibnamefont
			{Peotta}}\ and\ \bibinfo {author} {\bibfnamefont {P.}~\bibnamefont
			{T{\"o}rm{\"a}}},\ }\bibfield  {title} {\bibinfo {title} {{Superfluidity in
				topologically nontrivial flat bands}},\ }\href
	{https://www.nature.com/articles/ncomms9944} {\bibfield  {journal} {\bibinfo
			{journal} {Nat. Commun.}\ }\textbf {\bibinfo {volume} {6}},\ \bibinfo {pages}
		{8944} (\bibinfo {year} {2015})}\BibitemShut {NoStop}%
	\bibitem [{\citenamefont {Julku}\ \emph {et~al.}(2016)\citenamefont {Julku},
		\citenamefont {Peotta}, \citenamefont {Vanhala}, \citenamefont {Kim},\ and\
		\citenamefont {T\"orm\"a}}]{julku2016}%
	\BibitemOpen
	\bibfield  {author} {\bibinfo {author} {\bibfnamefont {A.}~\bibnamefont
			{Julku}}, \bibinfo {author} {\bibfnamefont {S.}~\bibnamefont {Peotta}},
		\bibinfo {author} {\bibfnamefont {T.~I.}\ \bibnamefont {Vanhala}}, \bibinfo
		{author} {\bibfnamefont {D.-H.}\ \bibnamefont {Kim}},\ and\ \bibinfo {author}
		{\bibfnamefont {P.}~\bibnamefont {T\"orm\"a}},\ }\bibfield  {title} {\bibinfo
		{title} {{Geometric Origin of Superfluidity in the Lieb-Lattice Flat Band}},\
	}\href {https://doi.org/10.1103/PhysRevLett.117.045303} {\bibfield  {journal}
		{\bibinfo  {journal} {Phys. Rev. Lett.}\ }\textbf {\bibinfo {volume} {117}},\
		\bibinfo {pages} {045303} (\bibinfo {year} {2016})}\BibitemShut {NoStop}%
	\bibitem [{\citenamefont {Liang}\ \emph {et~al.}(2017)\citenamefont {Liang},
		\citenamefont {Vanhala}, \citenamefont {Peotta}, \citenamefont {Siro},
		\citenamefont {Harju},\ and\ \citenamefont {T\"orm\"a}}]{lianglong2017}%
	\BibitemOpen
	\bibfield  {author} {\bibinfo {author} {\bibfnamefont {L.}~\bibnamefont
			{Liang}}, \bibinfo {author} {\bibfnamefont {T.~I.}\ \bibnamefont {Vanhala}},
		\bibinfo {author} {\bibfnamefont {S.}~\bibnamefont {Peotta}}, \bibinfo
		{author} {\bibfnamefont {T.}~\bibnamefont {Siro}}, \bibinfo {author}
		{\bibfnamefont {A.}~\bibnamefont {Harju}},\ and\ \bibinfo {author}
		{\bibfnamefont {P.}~\bibnamefont {T\"orm\"a}},\ }\bibfield  {title} {\bibinfo
		{title} {{Band geometry, Berry curvature, and superfluid weight}},\ }\href
	{https://doi.org/10.1103/PhysRevB.95.024515} {\bibfield  {journal} {\bibinfo
			{journal} {Phys. Rev. B}\ }\textbf {\bibinfo {volume} {95}},\ \bibinfo
		{pages} {024515} (\bibinfo {year} {2017})}\BibitemShut {NoStop}%
	\bibitem [{\citenamefont {T\"orm\"a}\ \emph {et~al.}(2018)\citenamefont
		{T\"orm\"a}, \citenamefont {Liang},\ and\ \citenamefont
		{Peotta}}]{torma2018}%
	\BibitemOpen
	\bibfield  {author} {\bibinfo {author} {\bibfnamefont {P.}~\bibnamefont
			{T\"orm\"a}}, \bibinfo {author} {\bibfnamefont {L.}~\bibnamefont {Liang}},\
		and\ \bibinfo {author} {\bibfnamefont {S.}~\bibnamefont {Peotta}},\
	}\bibfield  {title} {\bibinfo {title} {{Quantum metric and effective mass of
				a two-body bound state in a flat band}},\ }\href
	{https://doi.org/10.1103/PhysRevB.98.220511} {\bibfield  {journal} {\bibinfo
			{journal} {Phys. Rev. B}\ }\textbf {\bibinfo {volume} {98}},\ \bibinfo
		{pages} {220511} (\bibinfo {year} {2018})}\BibitemShut {NoStop}%
	\bibitem [{\citenamefont {Huhtinen}\ \emph {et~al.}(2022)\citenamefont
		{Huhtinen}, \citenamefont {Herzog-Arbeitman}, \citenamefont {Chew},
		\citenamefont {Bernevig},\ and\ \citenamefont {T\"orm\"a}}]{huhtinen2022}%
	\BibitemOpen
	\bibfield  {author} {\bibinfo {author} {\bibfnamefont {K.-E.}\ \bibnamefont
			{Huhtinen}}, \bibinfo {author} {\bibfnamefont {J.}~\bibnamefont
			{Herzog-Arbeitman}}, \bibinfo {author} {\bibfnamefont {A.}~\bibnamefont
			{Chew}}, \bibinfo {author} {\bibfnamefont {B.~A.}\ \bibnamefont {Bernevig}},\
		and\ \bibinfo {author} {\bibfnamefont {P.}~\bibnamefont {T\"orm\"a}},\
	}\bibfield  {title} {\bibinfo {title} {{Revisiting flat band
				superconductivity: Dependence on minimal quantum metric and band
				touchings}},\ }\href {https://doi.org/10.1103/PhysRevB.106.014518} {\bibfield
		{journal} {\bibinfo  {journal} {Phys. Rev. B}\ }\textbf {\bibinfo {volume}
			{106}},\ \bibinfo {pages} {014518} (\bibinfo {year} {2022})}\BibitemShut
	{NoStop}%
	\bibitem [{\citenamefont {Parameswaran}\ \emph {et~al.}(2012)\citenamefont
		{Parameswaran}, \citenamefont {Roy},\ and\ \citenamefont
		{Sondhi}}]{parameswaran2012}%
	\BibitemOpen
	\bibfield  {author} {\bibinfo {author} {\bibfnamefont {S.~A.}\ \bibnamefont
			{Parameswaran}}, \bibinfo {author} {\bibfnamefont {R.}~\bibnamefont {Roy}},\
		and\ \bibinfo {author} {\bibfnamefont {S.~L.}\ \bibnamefont {Sondhi}},\
	}\bibfield  {title} {\bibinfo {title} {{Fractional Chern insulators and the
				${W}_{\ensuremath{\infty}}$ algebra}},\ }\href
	{https://doi.org/10.1103/PhysRevB.85.241308} {\bibfield  {journal} {\bibinfo
			{journal} {Phys. Rev. B}\ }\textbf {\bibinfo {volume} {85}},\ \bibinfo
		{pages} {241308} (\bibinfo {year} {2012})}\BibitemShut {NoStop}%
	\bibitem [{\citenamefont {Roy}(2014)}]{roy2014}%
	\BibitemOpen
	\bibfield  {author} {\bibinfo {author} {\bibfnamefont {R.}~\bibnamefont
			{Roy}},\ }\bibfield  {title} {\bibinfo {title} {{Band geometry of fractional
				topological insulators}},\ }\href
	{https://doi.org/10.1103/PhysRevB.90.165139} {\bibfield  {journal} {\bibinfo
			{journal} {Phys. Rev. B}\ }\textbf {\bibinfo {volume} {90}},\ \bibinfo
		{pages} {165139} (\bibinfo {year} {2014})}\BibitemShut {NoStop}%
	\bibitem [{\citenamefont {Jackson}\ \emph {et~al.}(2015)\citenamefont
		{Jackson}, \citenamefont {M{\"o}ller},\ and\ \citenamefont
		{Roy}}]{jackson2015}%
	\BibitemOpen
	\bibfield  {author} {\bibinfo {author} {\bibfnamefont {T.~S.}\ \bibnamefont
			{Jackson}}, \bibinfo {author} {\bibfnamefont {G.}~\bibnamefont
			{M{\"o}ller}},\ and\ \bibinfo {author} {\bibfnamefont {R.}~\bibnamefont
			{Roy}},\ }\bibfield  {title} {\bibinfo {title} {{Geometric stability of
				topological lattice phases}},\ }\href
	{https://www.nature.com/articles/ncomms9629} {\bibfield  {journal} {\bibinfo
			{journal} {Nat. Commun.}\ }\textbf {\bibinfo {volume} {6}},\ \bibinfo {pages}
		{8629} (\bibinfo {year} {2015})}\BibitemShut {NoStop}%
	\bibitem [{\citenamefont {Gao}\ \emph {et~al.}(2014)\citenamefont {Gao},
		\citenamefont {Yang},\ and\ \citenamefont {Niu}}]{GaoY14prl}%
	\BibitemOpen
	\bibfield  {author} {\bibinfo {author} {\bibfnamefont {Y.}~\bibnamefont
			{Gao}}, \bibinfo {author} {\bibfnamefont {S.~A.}\ \bibnamefont {Yang}},\ and\
		\bibinfo {author} {\bibfnamefont {Q.}~\bibnamefont {Niu}},\ }\bibfield
	{title} {\bibinfo {title} {{Field Induced Positional Shift of Bloch Electrons
				and Its Dynamical Implications}},\ }\href
	{https://doi.org/10.1103/PhysRevLett.112.166601} {\bibfield  {journal}
		{\bibinfo  {journal} {Phys. Rev. Lett.}\ }\textbf {\bibinfo {volume} {112}},\
		\bibinfo {pages} {166601} (\bibinfo {year} {2014})}\BibitemShut {NoStop}%
	\bibitem [{\citenamefont {Sodemann}\ and\ \citenamefont {Fu}(2015)}]{FuL15prl}%
	\BibitemOpen
	\bibfield  {author} {\bibinfo {author} {\bibfnamefont {I.}~\bibnamefont
			{Sodemann}}\ and\ \bibinfo {author} {\bibfnamefont {L.}~\bibnamefont {Fu}},\
	}\bibfield  {title} {\bibinfo {title} {{Quantum Nonlinear Hall Effect Induced
				by Berry Curvature Dipole in Time-Reversal Invariant Materials}},\ }\href
	{https://doi.org/10.1103/PhysRevLett.115.216806} {\bibfield  {journal}
		{\bibinfo  {journal} {Phys. Rev. Lett.}\ }\textbf {\bibinfo {volume} {115}},\
		\bibinfo {pages} {216806} (\bibinfo {year} {2015})}\BibitemShut {NoStop}%
	\bibitem [{\citenamefont {Du}\ \emph {et~al.}(2018)\citenamefont {Du},
		\citenamefont {Wang}, \citenamefont {Lu},\ and\ \citenamefont
		{Xie}}]{Lu18prl}%
	\BibitemOpen
	\bibfield  {author} {\bibinfo {author} {\bibfnamefont {Z.~Z.}\ \bibnamefont
			{Du}}, \bibinfo {author} {\bibfnamefont {C.~M.}\ \bibnamefont {Wang}},
		\bibinfo {author} {\bibfnamefont {H.-Z.}\ \bibnamefont {Lu}},\ and\ \bibinfo
		{author} {\bibfnamefont {X.~C.}\ \bibnamefont {Xie}},\ }\bibfield  {title}
	{\bibinfo {title} {{Band Signatures for Strong Nonlinear Hall Effect in
				Bilayer ${\mathrm{WTe}}_{2}$}},\ }\href
	{https://doi.org/10.1103/PhysRevLett.121.266601} {\bibfield  {journal}
		{\bibinfo  {journal} {Phys. Rev. Lett.}\ }\textbf {\bibinfo {volume} {121}},\
		\bibinfo {pages} {266601} (\bibinfo {year} {2018})}\BibitemShut {NoStop}%
	\bibitem [{\citenamefont {Ma}\ \emph {et~al.}(2019)\citenamefont {Ma},
		\citenamefont {Xu}, \citenamefont {Shen}, \citenamefont {MacNeill},
		\citenamefont {Fatemi}, \citenamefont {Chang}, \citenamefont {Mier~Valdivia},
		\citenamefont {Wu}, \citenamefont {Du}, \citenamefont {Hsu} \emph
		{et~al.}}]{XuSY19nature}%
	\BibitemOpen
	\bibfield  {author} {\bibinfo {author} {\bibfnamefont {Q.}~\bibnamefont
			{Ma}}, \bibinfo {author} {\bibfnamefont {S.-Y.}\ \bibnamefont {Xu}}, \bibinfo
		{author} {\bibfnamefont {H.}~\bibnamefont {Shen}}, \bibinfo {author}
		{\bibfnamefont {D.}~\bibnamefont {MacNeill}}, \bibinfo {author}
		{\bibfnamefont {V.}~\bibnamefont {Fatemi}}, \bibinfo {author} {\bibfnamefont
			{T.-R.}\ \bibnamefont {Chang}}, \bibinfo {author} {\bibfnamefont {A.~M.}\
			\bibnamefont {Mier~Valdivia}}, \bibinfo {author} {\bibfnamefont
			{S.}~\bibnamefont {Wu}}, \bibinfo {author} {\bibfnamefont {Z.}~\bibnamefont
			{Du}}, \bibinfo {author} {\bibfnamefont {C.-H.}\ \bibnamefont {Hsu}}, \emph
		{et~al.},\ }\bibfield  {title} {\bibinfo {title} {{Observation of the
				nonlinear Hall effect under time-reversal-symmetric conditions}},\ }\href
	{https://doi.org/10.1038/s41586-018-0807-6} {\bibfield  {journal} {\bibinfo
			{journal} {Nature}\ }\textbf {\bibinfo {volume} {565}},\ \bibinfo {pages}
		{337} (\bibinfo {year} {2019})}\BibitemShut {NoStop}%
	\bibitem [{\citenamefont {Du}\ \emph {et~al.}(2019)\citenamefont {Du},
		\citenamefont {Wang}, \citenamefont {Li}, \citenamefont {Lu},\ and\
		\citenamefont {Xie}}]{Lu19nc}%
	\BibitemOpen
	\bibfield  {author} {\bibinfo {author} {\bibfnamefont {Z.}~\bibnamefont
			{Du}}, \bibinfo {author} {\bibfnamefont {C.}~\bibnamefont {Wang}}, \bibinfo
		{author} {\bibfnamefont {S.}~\bibnamefont {Li}}, \bibinfo {author}
		{\bibfnamefont {H.-Z.}\ \bibnamefont {Lu}},\ and\ \bibinfo {author}
		{\bibfnamefont {X.}~\bibnamefont {Xie}},\ }\bibfield  {title} {\bibinfo
		{title} {{Disorder-induced nonlinear Hall effect with time-reversal
				symmetry}},\ }\href {https://doi.org/10.1038/s41467-019-10941-3} {\bibfield
		{journal} {\bibinfo  {journal} {Nat. Commun.}\ }\textbf {\bibinfo {volume}
			{10}},\ \bibinfo {pages} {3047} (\bibinfo {year} {2019})}\BibitemShut
	{NoStop}%
	\bibitem [{\citenamefont {Wang}\ \emph {et~al.}(2021)\citenamefont {Wang},
		\citenamefont {Gao},\ and\ \citenamefont {Xiao}}]{GaoY21prl}%
	\BibitemOpen
	\bibfield  {author} {\bibinfo {author} {\bibfnamefont {C.}~\bibnamefont
			{Wang}}, \bibinfo {author} {\bibfnamefont {Y.}~\bibnamefont {Gao}},\ and\
		\bibinfo {author} {\bibfnamefont {D.}~\bibnamefont {Xiao}},\ }\bibfield
	{title} {\bibinfo {title} {{Intrinsic Nonlinear Hall Effect in
				Antiferromagnetic Tetragonal CuMnAs}},\ }\href
	{https://doi.org/10.1103/PhysRevLett.127.277201} {\bibfield  {journal}
		{\bibinfo  {journal} {Phys. Rev. Lett.}\ }\textbf {\bibinfo {volume} {127}},\
		\bibinfo {pages} {277201} (\bibinfo {year} {2021})}\BibitemShut {NoStop}%
	\bibitem [{\citenamefont {Liu}\ \emph {et~al.}(2021)\citenamefont {Liu},
		\citenamefont {Zhao}, \citenamefont {Huang}, \citenamefont {Wu},
		\citenamefont {Sheng}, \citenamefont {Xiao},\ and\ \citenamefont
		{Yang}}]{YangSY21prl}%
	\BibitemOpen
	\bibfield  {author} {\bibinfo {author} {\bibfnamefont {H.}~\bibnamefont
			{Liu}}, \bibinfo {author} {\bibfnamefont {J.}~\bibnamefont {Zhao}}, \bibinfo
		{author} {\bibfnamefont {Y.-X.}\ \bibnamefont {Huang}}, \bibinfo {author}
		{\bibfnamefont {W.}~\bibnamefont {Wu}}, \bibinfo {author} {\bibfnamefont
			{X.-L.}\ \bibnamefont {Sheng}}, \bibinfo {author} {\bibfnamefont
			{C.}~\bibnamefont {Xiao}},\ and\ \bibinfo {author} {\bibfnamefont {S.~A.}\
			\bibnamefont {Yang}},\ }\bibfield  {title} {\bibinfo {title} {{Intrinsic
				Second-Order Anomalous Hall Effect and Its Application in Compensated
				Antiferromagnets}},\ }\href {https://doi.org/10.1103/PhysRevLett.127.277202}
	{\bibfield  {journal} {\bibinfo  {journal} {Phys. Rev. Lett.}\ }\textbf
		{\bibinfo {volume} {127}},\ \bibinfo {pages} {277202} (\bibinfo {year}
		{2021})}\BibitemShut {NoStop}%
	\bibitem [{\citenamefont {Gao}\ \emph {et~al.}(2023)\citenamefont {Gao},
		\citenamefont {Liu}, \citenamefont {Qiu}, \citenamefont {Ghosh},
		\citenamefont {V.~Trevisan}, \citenamefont {Onishi}, \citenamefont {Hu},
		\citenamefont {Qian}, \citenamefont {Tien}, \citenamefont {Chen} \emph
		{et~al.}}]{XuSY23science}%
	\BibitemOpen
	\bibfield  {author} {\bibinfo {author} {\bibfnamefont {A.}~\bibnamefont
			{Gao}}, \bibinfo {author} {\bibfnamefont {Y.-F.}\ \bibnamefont {Liu}},
		\bibinfo {author} {\bibfnamefont {J.-X.}\ \bibnamefont {Qiu}}, \bibinfo
		{author} {\bibfnamefont {B.}~\bibnamefont {Ghosh}}, \bibinfo {author}
		{\bibfnamefont {T.}~\bibnamefont {V.~Trevisan}}, \bibinfo {author}
		{\bibfnamefont {Y.}~\bibnamefont {Onishi}}, \bibinfo {author} {\bibfnamefont
			{C.}~\bibnamefont {Hu}}, \bibinfo {author} {\bibfnamefont {T.}~\bibnamefont
			{Qian}}, \bibinfo {author} {\bibfnamefont {H.-J.}\ \bibnamefont {Tien}},
		\bibinfo {author} {\bibfnamefont {S.-W.}\ \bibnamefont {Chen}}, \emph
		{et~al.},\ }\bibfield  {title} {\bibinfo {title} {{Quantum metric nonlinear
				Hall effect in a topological antiferromagnetic heterostructure}},\ }\href
	{https://doi.org/10.1126/science.adf1506} {\bibfield  {journal} {\bibinfo
			{journal} {Science}\ }\textbf {\bibinfo {volume} {381}},\ \bibinfo {pages}
		{181} (\bibinfo {year} {2023})}\BibitemShut {NoStop}%
	\bibitem [{\citenamefont {Wang}\ \emph {et~al.}(2023)\citenamefont {Wang},
		\citenamefont {Kaplan}, \citenamefont {Zhang}, \citenamefont {Holder},
		\citenamefont {Cao}, \citenamefont {Wang}, \citenamefont {Zhou},
		\citenamefont {Zhou}, \citenamefont {Jiang}, \citenamefont {Zhang} \emph
		{et~al.}}]{GaoWB23nature}%
	\BibitemOpen
	\bibfield  {author} {\bibinfo {author} {\bibfnamefont {N.}~\bibnamefont
			{Wang}}, \bibinfo {author} {\bibfnamefont {D.}~\bibnamefont {Kaplan}},
		\bibinfo {author} {\bibfnamefont {Z.}~\bibnamefont {Zhang}}, \bibinfo
		{author} {\bibfnamefont {T.}~\bibnamefont {Holder}}, \bibinfo {author}
		{\bibfnamefont {N.}~\bibnamefont {Cao}}, \bibinfo {author} {\bibfnamefont
			{A.}~\bibnamefont {Wang}}, \bibinfo {author} {\bibfnamefont {X.}~\bibnamefont
			{Zhou}}, \bibinfo {author} {\bibfnamefont {F.}~\bibnamefont {Zhou}}, \bibinfo
		{author} {\bibfnamefont {Z.}~\bibnamefont {Jiang}}, \bibinfo {author}
		{\bibfnamefont {C.}~\bibnamefont {Zhang}}, \emph {et~al.},\ }\bibfield
	{title} {\bibinfo {title} {Quantum-metric-induced nonlinear transport in a
			topological antiferromagnet},\ }\href
	{https://doi.org/10.1038/s41586-023-06363-3} {\bibfield  {journal} {\bibinfo
			{journal} {Nature}\ }\textbf {\bibinfo {volume} {621}},\ \bibinfo {pages}
		{487} (\bibinfo {year} {2023})}\BibitemShut {NoStop}%
	\bibitem [{\citenamefont {Qiang}\ \emph {et~al.}(2025)\citenamefont {Qiang},
		\citenamefont {Liu}, \citenamefont {Gao}, \citenamefont {Lu},\ and\
		\citenamefont {Xie}}]{Lu25as}%
	\BibitemOpen
	\bibfield  {author} {\bibinfo {author} {\bibfnamefont {X.-B.}\ \bibnamefont
			{Qiang}}, \bibinfo {author} {\bibfnamefont {T.}~\bibnamefont {Liu}}, \bibinfo
		{author} {\bibfnamefont {Z.-X.}\ \bibnamefont {Gao}}, \bibinfo {author}
		{\bibfnamefont {H.-Z.}\ \bibnamefont {Lu}},\ and\ \bibinfo {author}
		{\bibfnamefont {X.~C.}\ \bibnamefont {Xie}},\ }\bibfield  {title} {\bibinfo
		{title} {{A Clarification on Quantum-Metric-Induced Nonlinear Transport}},\
	}\href {https://doi.org/10.1002/advs.202514818} {\bibfield  {journal}
		{\bibinfo  {journal} {Adv. Sci.}\ }\textbf {\bibinfo {volume} {13}},\
		\bibinfo {pages} {e14818} (\bibinfo {year} {2025})}\BibitemShut {NoStop}%
	\bibitem [{\citenamefont {Edelstein}(1990)}]{Edestein90ssc}%
	\BibitemOpen
	\bibfield  {author} {\bibinfo {author} {\bibfnamefont {V.}~\bibnamefont
			{Edelstein}},\ }\bibfield  {title} {\bibinfo {title} {Spin polarization of
			conduction electrons induced by electric current in two-dimensional
			asymmetric electron systems},\ }\href
	{https://doi.org/https://doi.org/10.1016/0038-1098(90)90963-C} {\bibfield
		{journal} {\bibinfo  {journal} {Solid State Commun.}\ }\textbf {\bibinfo
			{volume} {73}},\ \bibinfo {pages} {233} (\bibinfo {year} {1990})}\BibitemShut
	{NoStop}%
	\bibitem [{\citenamefont {Kato}\ \emph
		{et~al.}(2004{\natexlab{a}})\citenamefont {Kato}, \citenamefont {Myers},
		\citenamefont {Gossard},\ and\ \citenamefont {Awschalom}}]{Kato04prl}%
	\BibitemOpen
	\bibfield  {author} {\bibinfo {author} {\bibfnamefont {Y.~K.}\ \bibnamefont
			{Kato}}, \bibinfo {author} {\bibfnamefont {R.~C.}\ \bibnamefont {Myers}},
		\bibinfo {author} {\bibfnamefont {A.~C.}\ \bibnamefont {Gossard}},\ and\
		\bibinfo {author} {\bibfnamefont {D.~D.}\ \bibnamefont {Awschalom}},\
	}\bibfield  {title} {\bibinfo {title} {{Current-Induced Spin Polarization in
				Strained Semiconductors}},\ }\href
	{https://doi.org/10.1103/PhysRevLett.93.176601} {\bibfield  {journal}
		{\bibinfo  {journal} {Phys. Rev. Lett.}\ }\textbf {\bibinfo {volume} {93}},\
		\bibinfo {pages} {176601} (\bibinfo {year} {2004}{\natexlab{a}})}\BibitemShut
	{NoStop}%
	\bibitem [{\citenamefont {Bernevig}\ \emph {et~al.}(2005)\citenamefont
		{Bernevig}, \citenamefont {Hughes},\ and\ \citenamefont
		{Zhang}}]{ZhangSC05prl}%
	\BibitemOpen
	\bibfield  {author} {\bibinfo {author} {\bibfnamefont {B.~A.}\ \bibnamefont
			{Bernevig}}, \bibinfo {author} {\bibfnamefont {T.~L.}\ \bibnamefont
			{Hughes}},\ and\ \bibinfo {author} {\bibfnamefont {S.-C.}\ \bibnamefont
			{Zhang}},\ }\bibfield  {title} {\bibinfo {title} {{Orbitronics: The Intrinsic
				Orbital Current in $p$-Doped Silicon}},\ }\href
	{https://doi.org/10.1103/PhysRevLett.95.066601} {\bibfield  {journal}
		{\bibinfo  {journal} {Phys. Rev. Lett.}\ }\textbf {\bibinfo {volume} {95}},\
		\bibinfo {pages} {066601} (\bibinfo {year} {2005})}\BibitemShut {NoStop}%
	\bibitem [{\citenamefont {Tanaka}\ \emph {et~al.}(2008)\citenamefont {Tanaka},
		\citenamefont {Kontani}, \citenamefont {Naito}, \citenamefont {Naito},
		\citenamefont {Hirashima}, \citenamefont {Yamada},\ and\ \citenamefont
		{Inoue}}]{Inoue08prb}%
	\BibitemOpen
	\bibfield  {author} {\bibinfo {author} {\bibfnamefont {T.}~\bibnamefont
			{Tanaka}}, \bibinfo {author} {\bibfnamefont {H.}~\bibnamefont {Kontani}},
		\bibinfo {author} {\bibfnamefont {M.}~\bibnamefont {Naito}}, \bibinfo
		{author} {\bibfnamefont {T.}~\bibnamefont {Naito}}, \bibinfo {author}
		{\bibfnamefont {D.~S.}\ \bibnamefont {Hirashima}}, \bibinfo {author}
		{\bibfnamefont {K.}~\bibnamefont {Yamada}},\ and\ \bibinfo {author}
		{\bibfnamefont {J.}~\bibnamefont {Inoue}},\ }\bibfield  {title} {\bibinfo
		{title} {{Intrinsic spin Hall effect and orbital Hall effect in $4d$ and $5d$
				transition metals}},\ }\href {https://doi.org/10.1103/PhysRevB.77.165117}
	{\bibfield  {journal} {\bibinfo  {journal} {Phys. Rev. B}\ }\textbf {\bibinfo
			{volume} {77}},\ \bibinfo {pages} {165117} (\bibinfo {year}
		{2008})}\BibitemShut {NoStop}%
	\bibitem [{\citenamefont {Chernyshov}\ \emph {et~al.}(2009)\citenamefont
		{Chernyshov}, \citenamefont {Overby}, \citenamefont {Liu}, \citenamefont
		{Furdyna}, \citenamefont {Lyanda-Geller},\ and\ \citenamefont
		{Rokhinson}}]{Leonid09np}%
	\BibitemOpen
	\bibfield  {author} {\bibinfo {author} {\bibfnamefont {A.}~\bibnamefont
			{Chernyshov}}, \bibinfo {author} {\bibfnamefont {M.}~\bibnamefont {Overby}},
		\bibinfo {author} {\bibfnamefont {X.}~\bibnamefont {Liu}}, \bibinfo {author}
		{\bibfnamefont {J.~K.}\ \bibnamefont {Furdyna}}, \bibinfo {author}
		{\bibfnamefont {Y.}~\bibnamefont {Lyanda-Geller}},\ and\ \bibinfo {author}
		{\bibfnamefont {L.~P.}\ \bibnamefont {Rokhinson}},\ }\bibfield  {title}
	{\bibinfo {title} {Evidence for reversible control of magnetization in a
			ferromagnetic material by means of spin--orbit magnetic field},\ }\href
	{https://doi.org/10.1038/nphys1362} {\bibfield  {journal} {\bibinfo
			{journal} {Nat. Phys.}\ }\textbf {\bibinfo {volume} {5}},\ \bibinfo {pages}
		{656} (\bibinfo {year} {2009})}\BibitemShut {NoStop}%
	\bibitem [{\citenamefont {Fang}\ \emph {et~al.}(2011)\citenamefont {Fang},
		\citenamefont {Kurebayashi}, \citenamefont {Wunderlich}, \citenamefont
		{V{\`y}born{\`y}}, \citenamefont {Z{\^a}rbo}, \citenamefont {Campion},
		\citenamefont {Casiraghi}, \citenamefont {Gallagher}, \citenamefont
		{Jungwirth},\ and\ \citenamefont {Ferguson}}]{Ferguson11nn}%
	\BibitemOpen
	\bibfield  {author} {\bibinfo {author} {\bibfnamefont {D.}~\bibnamefont
			{Fang}}, \bibinfo {author} {\bibfnamefont {H.}~\bibnamefont {Kurebayashi}},
		\bibinfo {author} {\bibfnamefont {J.}~\bibnamefont {Wunderlich}}, \bibinfo
		{author} {\bibfnamefont {K.}~\bibnamefont {V{\`y}born{\`y}}}, \bibinfo
		{author} {\bibfnamefont {L.~P.}\ \bibnamefont {Z{\^a}rbo}}, \bibinfo {author}
		{\bibfnamefont {R.}~\bibnamefont {Campion}}, \bibinfo {author} {\bibfnamefont
			{A.}~\bibnamefont {Casiraghi}}, \bibinfo {author} {\bibfnamefont
			{B.}~\bibnamefont {Gallagher}}, \bibinfo {author} {\bibfnamefont
			{T.}~\bibnamefont {Jungwirth}},\ and\ \bibinfo {author} {\bibfnamefont
			{A.}~\bibnamefont {Ferguson}},\ }\bibfield  {title} {\bibinfo {title}
		{Spin--orbit-driven ferromagnetic resonance},\ }\href
	{https://doi.org/10.1038/nnano.2011.68} {\bibfield  {journal} {\bibinfo
			{journal} {Nat. Nanotechnol.}\ }\textbf {\bibinfo {volume} {6}},\ \bibinfo
		{pages} {413} (\bibinfo {year} {2011})}\BibitemShut {NoStop}%
	\bibitem [{\citenamefont {Culcer}\ and\ \citenamefont
		{Winkler}(2007)}]{Dimi07prl}%
	\BibitemOpen
	\bibfield  {author} {\bibinfo {author} {\bibfnamefont {D.}~\bibnamefont
			{Culcer}}\ and\ \bibinfo {author} {\bibfnamefont {R.}~\bibnamefont
			{Winkler}},\ }\bibfield  {title} {\bibinfo {title} {{Generation of Spin
				Currents and Spin Densities in Systems with Reduced Symmetry}},\ }\href
	{https://doi.org/10.1103/PhysRevLett.99.226601} {\bibfield  {journal}
		{\bibinfo  {journal} {Phys. Rev. Lett.}\ }\textbf {\bibinfo {volume} {99}},\
		\bibinfo {pages} {226601} (\bibinfo {year} {2007})}\BibitemShut {NoStop}%
	\bibitem [{\citenamefont {\ifmmode~\check{Z}\else \v{Z}\fi{}elezn\'y}\ \emph
		{et~al.}(2014)\citenamefont {\ifmmode~\check{Z}\else \v{Z}\fi{}elezn\'y},
		\citenamefont {Gao}, \citenamefont {V\'yborn\'y}, \citenamefont {Zemen},
		\citenamefont {Ma\ifmmode~\check{s}\else \v{s}\fi{}ek}, \citenamefont
		{Manchon}, \citenamefont {Wunderlich}, \citenamefont {Sinova},\ and\
		\citenamefont {Jungwirth}}]{Sinowa14prl}%
	\BibitemOpen
	\bibfield  {author} {\bibinfo {author} {\bibfnamefont {J.}~\bibnamefont
			{\ifmmode~\check{Z}\else \v{Z}\fi{}elezn\'y}}, \bibinfo {author}
		{\bibfnamefont {H.}~\bibnamefont {Gao}}, \bibinfo {author} {\bibfnamefont
			{K.}~\bibnamefont {V\'yborn\'y}}, \bibinfo {author} {\bibfnamefont
			{J.}~\bibnamefont {Zemen}}, \bibinfo {author} {\bibfnamefont
			{J.}~\bibnamefont {Ma\ifmmode~\check{s}\else \v{s}\fi{}ek}}, \bibinfo
		{author} {\bibfnamefont {A.}~\bibnamefont {Manchon}}, \bibinfo {author}
		{\bibfnamefont {J.}~\bibnamefont {Wunderlich}}, \bibinfo {author}
		{\bibfnamefont {J.}~\bibnamefont {Sinova}},\ and\ \bibinfo {author}
		{\bibfnamefont {T.}~\bibnamefont {Jungwirth}},\ }\bibfield  {title} {\bibinfo
		{title} {{Relativistic N\'eel-Order Fields Induced by Electrical Current in
				Antiferromagnets}},\ }\href {https://doi.org/10.1103/PhysRevLett.113.157201}
	{\bibfield  {journal} {\bibinfo  {journal} {Phys. Rev. Lett.}\ }\textbf
		{\bibinfo {volume} {113}},\ \bibinfo {pages} {157201} (\bibinfo {year}
		{2014})}\BibitemShut {NoStop}%
	\bibitem [{\citenamefont {Manchon}\ \emph {et~al.}(2019)\citenamefont
		{Manchon}, \citenamefont {\ifmmode~\check{Z}\else \v{Z}\fi{}elezn\'y},
		\citenamefont {Miron}, \citenamefont {Jungwirth}, \citenamefont {Sinova},
		\citenamefont {Thiaville}, \citenamefont {Garello},\ and\ \citenamefont
		{Gambardella}}]{Manchon19rmp}%
	\BibitemOpen
	\bibfield  {author} {\bibinfo {author} {\bibfnamefont {A.}~\bibnamefont
			{Manchon}}, \bibinfo {author} {\bibfnamefont {J.}~\bibnamefont
			{\ifmmode~\check{Z}\else \v{Z}\fi{}elezn\'y}}, \bibinfo {author}
		{\bibfnamefont {I.~M.}\ \bibnamefont {Miron}}, \bibinfo {author}
		{\bibfnamefont {T.}~\bibnamefont {Jungwirth}}, \bibinfo {author}
		{\bibfnamefont {J.}~\bibnamefont {Sinova}}, \bibinfo {author} {\bibfnamefont
			{A.}~\bibnamefont {Thiaville}}, \bibinfo {author} {\bibfnamefont
			{K.}~\bibnamefont {Garello}},\ and\ \bibinfo {author} {\bibfnamefont
			{P.}~\bibnamefont {Gambardella}},\ }\bibfield  {title} {\bibinfo {title}
		{Current-induced spin-orbit torques in ferromagnetic and antiferromagnetic
			systems},\ }\href {https://doi.org/10.1103/RevModPhys.91.035004} {\bibfield
		{journal} {\bibinfo  {journal} {Rev. Mod. Phys.}\ }\textbf {\bibinfo {volume}
			{91}},\ \bibinfo {pages} {035004} (\bibinfo {year} {2019})}\BibitemShut
	{NoStop}%
	\bibitem [{\citenamefont {Go}\ \emph {et~al.}(2021)\citenamefont {Go},
		\citenamefont {Jo}, \citenamefont {Lee}, \citenamefont {Kl{\"a}ui},\ and\
		\citenamefont {Mokrousov}}]{Yuriy21epl}%
	\BibitemOpen
	\bibfield  {author} {\bibinfo {author} {\bibfnamefont {D.}~\bibnamefont
			{Go}}, \bibinfo {author} {\bibfnamefont {D.}~\bibnamefont {Jo}}, \bibinfo
		{author} {\bibfnamefont {H.-W.}\ \bibnamefont {Lee}}, \bibinfo {author}
		{\bibfnamefont {M.}~\bibnamefont {Kl{\"a}ui}},\ and\ \bibinfo {author}
		{\bibfnamefont {Y.}~\bibnamefont {Mokrousov}},\ }\bibfield  {title} {\bibinfo
		{title} {{Orbitronics: Orbital currents in solids}},\ }\href
	{https://doi.org/10.1209/0295-5075/ac2653} {\bibfield  {journal} {\bibinfo
			{journal} {Europhys. Lett.}\ }\textbf {\bibinfo {volume} {135}},\ \bibinfo
		{pages} {37001} (\bibinfo {year} {2021})}\BibitemShut {NoStop}%
	\bibitem [{\citenamefont {Fert}\ \emph {et~al.}(2024)\citenamefont {Fert},
		\citenamefont {Ramesh}, \citenamefont {Garcia}, \citenamefont {Casanova},\
		and\ \citenamefont {Bibes}}]{Felix24rmp}%
	\BibitemOpen
	\bibfield  {author} {\bibinfo {author} {\bibfnamefont {A.}~\bibnamefont
			{Fert}}, \bibinfo {author} {\bibfnamefont {R.}~\bibnamefont {Ramesh}},
		\bibinfo {author} {\bibfnamefont {V.}~\bibnamefont {Garcia}}, \bibinfo
		{author} {\bibfnamefont {F.}~\bibnamefont {Casanova}},\ and\ \bibinfo
		{author} {\bibfnamefont {M.}~\bibnamefont {Bibes}},\ }\bibfield  {title}
	{\bibinfo {title} {Electrical control of magnetism by electric field and
			current-induced torques},\ }\href
	{https://doi.org/10.1103/RevModPhys.96.015005} {\bibfield  {journal}
		{\bibinfo  {journal} {Rev. Mod. Phys.}\ }\textbf {\bibinfo {volume} {96}},\
		\bibinfo {pages} {015005} (\bibinfo {year} {2024})}\BibitemShut {NoStop}%
	\bibitem [{\citenamefont {Jo}\ \emph {et~al.}(2024)\citenamefont {Jo},
		\citenamefont {Go}, \citenamefont {Choi},\ and\ \citenamefont
		{Lee}}]{Jo24npj}%
	\BibitemOpen
	\bibfield  {author} {\bibinfo {author} {\bibfnamefont {D.}~\bibnamefont
			{Jo}}, \bibinfo {author} {\bibfnamefont {D.}~\bibnamefont {Go}}, \bibinfo
		{author} {\bibfnamefont {G.-M.}\ \bibnamefont {Choi}},\ and\ \bibinfo
		{author} {\bibfnamefont {H.-W.}\ \bibnamefont {Lee}},\ }\bibfield  {title}
	{\bibinfo {title} {{Spintronics meets orbitronics: Emergence of orbital
				angular momentum in solids}},\ }\href
	{https://doi.org/10.1038/s44306-024-00023-6} {\bibfield  {journal} {\bibinfo
			{journal} {npj Spintronics}\ }\textbf {\bibinfo {volume} {2}},\ \bibinfo
		{pages} {19} (\bibinfo {year} {2024})}\BibitemShut {NoStop}%
	\bibitem [{\citenamefont {Wang}\ \emph
		{et~al.}(2024{\natexlab{a}})\citenamefont {Wang}, \citenamefont {Chen},
		\citenamefont {Yang}, \citenamefont {Hu}, \citenamefont {Li}, \citenamefont
		{Wang}, \citenamefont {Zhang},\ and\ \citenamefont {Jiang}}]{WangP24aem}%
	\BibitemOpen
	\bibfield  {author} {\bibinfo {author} {\bibfnamefont {P.}~\bibnamefont
			{Wang}}, \bibinfo {author} {\bibfnamefont {F.}~\bibnamefont {Chen}}, \bibinfo
		{author} {\bibfnamefont {Y.}~\bibnamefont {Yang}}, \bibinfo {author}
		{\bibfnamefont {S.}~\bibnamefont {Hu}}, \bibinfo {author} {\bibfnamefont
			{Y.}~\bibnamefont {Li}}, \bibinfo {author} {\bibfnamefont {W.}~\bibnamefont
			{Wang}}, \bibinfo {author} {\bibfnamefont {D.}~\bibnamefont {Zhang}},\ and\
		\bibinfo {author} {\bibfnamefont {Y.}~\bibnamefont {Jiang}},\ }\bibfield
	{title} {\bibinfo {title} {{Orbitronics: Mechanisms, Materials and
				Devices}},\ }\href {https://doi.org/10.1002/aelm.202400554} {\bibfield
		{journal} {\bibinfo  {journal} {Adv. Electron. Mater.}\ }\textbf {\bibinfo
			{volume} {11}},\ \bibinfo {pages} {2400554} (\bibinfo {year}
		{2024}{\natexlab{a}})}\BibitemShut {NoStop}%
	\bibitem [{\citenamefont {Qiang}\ \emph {et~al.}(2026)\citenamefont {Qiang},
		\citenamefont {Liu}, \citenamefont {Lu},\ and\ \citenamefont
		{Xie}}]{Lu26apl}%
	\BibitemOpen
	\bibfield  {author} {\bibinfo {author} {\bibfnamefont {X.-B.}\ \bibnamefont
			{Qiang}}, \bibinfo {author} {\bibfnamefont {T.}~\bibnamefont {Liu}}, \bibinfo
		{author} {\bibfnamefont {H.-Z.}\ \bibnamefont {Lu}},\ and\ \bibinfo {author}
		{\bibfnamefont {X.~C.}\ \bibnamefont {Xie}},\ }\bibfield  {title} {\bibinfo
		{title} {{Quantum geometric origin of orbital magnetization}},\ }\href
	{https://doi.org/10.1063/5.0310234} {\bibfield  {journal} {\bibinfo
			{journal} {Appl. Phys. Lett.}\ }\textbf {\bibinfo {volume} {128}},\ \bibinfo
		{pages} {010501} (\bibinfo {year} {2026})}\BibitemShut {NoStop}%
	\bibitem [{\citenamefont {Xiao}\ \emph {et~al.}(2005)\citenamefont {Xiao},
		\citenamefont {Shi},\ and\ \citenamefont {Niu}}]{Niu05prl}%
	\BibitemOpen
	\bibfield  {author} {\bibinfo {author} {\bibfnamefont {D.}~\bibnamefont
			{Xiao}}, \bibinfo {author} {\bibfnamefont {J.}~\bibnamefont {Shi}},\ and\
		\bibinfo {author} {\bibfnamefont {Q.}~\bibnamefont {Niu}},\ }\bibfield
	{title} {\bibinfo {title} {{Berry Phase Correction to Electron Density of
				States in Solids}},\ }\href {https://doi.org/10.1103/PhysRevLett.95.137204}
	{\bibfield  {journal} {\bibinfo  {journal} {Phys. Rev. Lett.}\ }\textbf
		{\bibinfo {volume} {95}},\ \bibinfo {pages} {137204} (\bibinfo {year}
		{2005})}\BibitemShut {NoStop}%
	\bibitem [{\citenamefont {Thonhauser}\ \emph {et~al.}(2005)\citenamefont
		{Thonhauser}, \citenamefont {Ceresoli}, \citenamefont {Vanderbilt},\ and\
		\citenamefont {Resta}}]{Resta05prl}%
	\BibitemOpen
	\bibfield  {author} {\bibinfo {author} {\bibfnamefont {T.}~\bibnamefont
			{Thonhauser}}, \bibinfo {author} {\bibfnamefont {D.}~\bibnamefont
			{Ceresoli}}, \bibinfo {author} {\bibfnamefont {D.}~\bibnamefont
			{Vanderbilt}},\ and\ \bibinfo {author} {\bibfnamefont {R.}~\bibnamefont
			{Resta}},\ }\bibfield  {title} {\bibinfo {title} {{Orbital Magnetization in
				Periodic Insulators}},\ }\href
	{https://doi.org/10.1103/PhysRevLett.95.137205} {\bibfield  {journal}
		{\bibinfo  {journal} {Phys. Rev. Lett.}\ }\textbf {\bibinfo {volume} {95}},\
		\bibinfo {pages} {137205} (\bibinfo {year} {2005})}\BibitemShut {NoStop}%
	\bibitem [{\citenamefont {Ceresoli}\ \emph {et~al.}(2006)\citenamefont
		{Ceresoli}, \citenamefont {Thonhauser}, \citenamefont {Vanderbilt},\ and\
		\citenamefont {Resta}}]{Resta06prb}%
	\BibitemOpen
	\bibfield  {author} {\bibinfo {author} {\bibfnamefont {D.}~\bibnamefont
			{Ceresoli}}, \bibinfo {author} {\bibfnamefont {T.}~\bibnamefont
			{Thonhauser}}, \bibinfo {author} {\bibfnamefont {D.}~\bibnamefont
			{Vanderbilt}},\ and\ \bibinfo {author} {\bibfnamefont {R.}~\bibnamefont
			{Resta}},\ }\bibfield  {title} {\bibinfo {title} {{Orbital magnetization in
				crystalline solids: Multi-band insulators, Chern insulators, and metals}},\
	}\href {https://doi.org/10.1103/PhysRevB.74.024408} {\bibfield  {journal}
		{\bibinfo  {journal} {Phys. Rev. B}\ }\textbf {\bibinfo {volume} {74}},\
		\bibinfo {pages} {024408} (\bibinfo {year} {2006})}\BibitemShut {NoStop}%
	\bibitem [{\citenamefont {Shi}\ \emph {et~al.}(2007)\citenamefont {Shi},
		\citenamefont {Vignale}, \citenamefont {Xiao},\ and\ \citenamefont
		{Niu}}]{Niu07prl}%
	\BibitemOpen
	\bibfield  {author} {\bibinfo {author} {\bibfnamefont {J.}~\bibnamefont
			{Shi}}, \bibinfo {author} {\bibfnamefont {G.}~\bibnamefont {Vignale}},
		\bibinfo {author} {\bibfnamefont {D.}~\bibnamefont {Xiao}},\ and\ \bibinfo
		{author} {\bibfnamefont {Q.}~\bibnamefont {Niu}},\ }\bibfield  {title}
	{\bibinfo {title} {{Quantum Theory of Orbital Magnetization and Its
				Generalization to Interacting Systems}},\ }\href
	{https://doi.org/10.1103/PhysRevLett.99.197202} {\bibfield  {journal}
		{\bibinfo  {journal} {Phys. Rev. Lett.}\ }\textbf {\bibinfo {volume} {99}},\
		\bibinfo {pages} {197202} (\bibinfo {year} {2007})}\BibitemShut {NoStop}%
	\bibitem [{\citenamefont {Xiao}\ \emph {et~al.}(2010)\citenamefont {Xiao},
		\citenamefont {Chang},\ and\ \citenamefont {Niu}}]{Xiao10rmp}%
	\BibitemOpen
	\bibfield  {author} {\bibinfo {author} {\bibfnamefont {D.}~\bibnamefont
			{Xiao}}, \bibinfo {author} {\bibfnamefont {M.-C.}\ \bibnamefont {Chang}},\
		and\ \bibinfo {author} {\bibfnamefont {Q.}~\bibnamefont {Niu}},\ }\bibfield
	{title} {\bibinfo {title} {Berry phase effects on electronic properties},\
	}\href {https://doi.org/10.1103/RevModPhys.82.1959} {\bibfield  {journal}
		{\bibinfo  {journal} {Rev. Mod. Phys.}\ }\textbf {\bibinfo {volume} {82}},\
		\bibinfo {pages} {1959} (\bibinfo {year} {2010})}\BibitemShut {NoStop}%
	\bibitem [{\citenamefont {Yoda}\ \emph {et~al.}(2015)\citenamefont {Yoda},
		\citenamefont {Yokoyama},\ and\ \citenamefont {Murakami}}]{Yoda15sr}%
	\BibitemOpen
	\bibfield  {author} {\bibinfo {author} {\bibfnamefont {T.}~\bibnamefont
			{Yoda}}, \bibinfo {author} {\bibfnamefont {T.}~\bibnamefont {Yokoyama}},\
		and\ \bibinfo {author} {\bibfnamefont {S.}~\bibnamefont {Murakami}},\
	}\bibfield  {title} {\bibinfo {title} {Current-induced orbital and spin
			magnetizations in crystals with helical structure},\ }\href
	{https://doi.org/10.1038/srep12024} {\bibfield  {journal} {\bibinfo
			{journal} {Sci. Rep.}\ }\textbf {\bibinfo {volume} {5}},\ \bibinfo {pages}
		{12024} (\bibinfo {year} {2015})}\BibitemShut {NoStop}%
	\bibitem [{\citenamefont {Zhong}\ \emph {et~al.}(2016)\citenamefont {Zhong},
		\citenamefont {Moore},\ and\ \citenamefont {Souza}}]{Moore16prl}%
	\BibitemOpen
	\bibfield  {author} {\bibinfo {author} {\bibfnamefont {S.}~\bibnamefont
			{Zhong}}, \bibinfo {author} {\bibfnamefont {J.~E.}\ \bibnamefont {Moore}},\
		and\ \bibinfo {author} {\bibfnamefont {I.}~\bibnamefont {Souza}},\ }\bibfield
	{title} {\bibinfo {title} {{Gyrotropic Magnetic Effect and the Magnetic
				Moment on the Fermi Surface}},\ }\href
	{https://doi.org/10.1103/PhysRevLett.116.077201} {\bibfield  {journal}
		{\bibinfo  {journal} {Phys. Rev. Lett.}\ }\textbf {\bibinfo {volume} {116}},\
		\bibinfo {pages} {077201} (\bibinfo {year} {2016})}\BibitemShut {NoStop}%
	\bibitem [{\citenamefont {Rou}\ \emph {et~al.}(2017)\citenamefont {Rou},
		\citenamefont {\ifmmode~\mbox{\c{S}}\else \c{S}\fi{}ahin}, \citenamefont
		{Ma},\ and\ \citenamefont {Pesin}}]{Pesin17prb}%
	\BibitemOpen
	\bibfield  {author} {\bibinfo {author} {\bibfnamefont {J.}~\bibnamefont
			{Rou}}, \bibinfo {author} {\bibfnamefont {C.}~\bibnamefont
			{\ifmmode~\mbox{\c{S}}\else \c{S}\fi{}ahin}}, \bibinfo {author}
		{\bibfnamefont {J.}~\bibnamefont {Ma}},\ and\ \bibinfo {author}
		{\bibfnamefont {D.~A.}\ \bibnamefont {Pesin}},\ }\bibfield  {title} {\bibinfo
		{title} {Kinetic orbital moments and nonlocal transport in disordered metals
			with nontrivial band geometry},\ }\href
	{https://doi.org/10.1103/PhysRevB.96.035120} {\bibfield  {journal} {\bibinfo
			{journal} {Phys. Rev. B}\ }\textbf {\bibinfo {volume} {96}},\ \bibinfo
		{pages} {035120} (\bibinfo {year} {2017})}\BibitemShut {NoStop}%
	\bibitem [{\citenamefont {Son}\ \emph {et~al.}(2019)\citenamefont {Son},
		\citenamefont {Kim}, \citenamefont {Ahn}, \citenamefont {Lee},\ and\
		\citenamefont {Lee}}]{Lee19prl}%
	\BibitemOpen
	\bibfield  {author} {\bibinfo {author} {\bibfnamefont {J.}~\bibnamefont
			{Son}}, \bibinfo {author} {\bibfnamefont {K.-H.}\ \bibnamefont {Kim}},
		\bibinfo {author} {\bibfnamefont {Y.~H.}\ \bibnamefont {Ahn}}, \bibinfo
		{author} {\bibfnamefont {H.-W.}\ \bibnamefont {Lee}},\ and\ \bibinfo {author}
		{\bibfnamefont {J.}~\bibnamefont {Lee}},\ }\bibfield  {title} {\bibinfo
		{title} {{Strain Engineering of the Berry Curvature Dipole and Valley
				Magnetization in Monolayer MoS$_2$}},\ }\href
	{https://doi.org/10.1103/PhysRevLett.123.036806} {\bibfield  {journal}
		{\bibinfo  {journal} {Phys. Rev. Lett.}\ }\textbf {\bibinfo {volume} {123}},\
		\bibinfo {pages} {036806} (\bibinfo {year} {2019})}\BibitemShut {NoStop}%
	\bibitem [{\citenamefont {Sharpe}\ \emph {et~al.}(2019)\citenamefont {Sharpe},
		\citenamefont {Fox}, \citenamefont {Barnard}, \citenamefont {Finney},
		\citenamefont {Watanabe}, \citenamefont {Taniguchi}, \citenamefont
		{Kastner},\ and\ \citenamefont {Goldhaber-Gordon}}]{Gorden19science}%
	\BibitemOpen
	\bibfield  {author} {\bibinfo {author} {\bibfnamefont {A.~L.}\ \bibnamefont
			{Sharpe}}, \bibinfo {author} {\bibfnamefont {E.~J.}\ \bibnamefont {Fox}},
		\bibinfo {author} {\bibfnamefont {A.~W.}\ \bibnamefont {Barnard}}, \bibinfo
		{author} {\bibfnamefont {J.}~\bibnamefont {Finney}}, \bibinfo {author}
		{\bibfnamefont {K.}~\bibnamefont {Watanabe}}, \bibinfo {author}
		{\bibfnamefont {T.}~\bibnamefont {Taniguchi}}, \bibinfo {author}
		{\bibfnamefont {M.}~\bibnamefont {Kastner}},\ and\ \bibinfo {author}
		{\bibfnamefont {D.}~\bibnamefont {Goldhaber-Gordon}},\ }\bibfield  {title}
	{\bibinfo {title} {{Emergent ferromagnetism near three-quarters filling in
				twisted bilayer graphene}},\ }\href@noop {} {\bibfield  {journal} {\bibinfo
			{journal} {Science}\ }\textbf {\bibinfo {volume} {365}},\ \bibinfo {pages}
		{605} (\bibinfo {year} {2019})}\BibitemShut {NoStop}%
	\bibitem [{\citenamefont {Hara}\ \emph {et~al.}(2020)\citenamefont {Hara},
		\citenamefont {Bahramy},\ and\ \citenamefont {Murakami}}]{Murakami20prb}%
	\BibitemOpen
	\bibfield  {author} {\bibinfo {author} {\bibfnamefont {D.}~\bibnamefont
			{Hara}}, \bibinfo {author} {\bibfnamefont {M.~S.}\ \bibnamefont {Bahramy}},\
		and\ \bibinfo {author} {\bibfnamefont {S.}~\bibnamefont {Murakami}},\
	}\bibfield  {title} {\bibinfo {title} {{Current-induced orbital magnetization
				in systems without inversion symmetry}},\ }\href
	{https://doi.org/10.1103/PhysRevB.102.184404} {\bibfield  {journal} {\bibinfo
			{journal} {Phys. Rev. B}\ }\textbf {\bibinfo {volume} {102}},\ \bibinfo
		{pages} {184404} (\bibinfo {year} {2020})}\BibitemShut {NoStop}%
	\bibitem [{\citenamefont {Johansson}(2024)}]{Johansson24jpcm}%
	\BibitemOpen
	\bibfield  {author} {\bibinfo {author} {\bibfnamefont {A.}~\bibnamefont
			{Johansson}},\ }\bibfield  {title} {\bibinfo {title} {{Theory of spin and
				orbital Edelstein effects}},\ }\href
	{https://dx.doi.org/10.1088/1361-648X/ad5e2b} {\bibfield  {journal} {\bibinfo
			{journal} {J. Phys.: Condens. Matter}\ }\textbf {\bibinfo {volume} {36}},\
		\bibinfo {pages} {423002} (\bibinfo {year} {2024})}\BibitemShut {NoStop}%
	\bibitem [{\citenamefont {Allwood}\ \emph {et~al.}(2003)\citenamefont
		{Allwood}, \citenamefont {Xiong}, \citenamefont {Cooke},\ and\ \citenamefont
		{Cowburn}}]{Cowburn03jpd}%
	\BibitemOpen
	\bibfield  {author} {\bibinfo {author} {\bibfnamefont {D.}~\bibnamefont
			{Allwood}}, \bibinfo {author} {\bibfnamefont {G.}~\bibnamefont {Xiong}},
		\bibinfo {author} {\bibfnamefont {M.}~\bibnamefont {Cooke}},\ and\ \bibinfo
		{author} {\bibfnamefont {R.}~\bibnamefont {Cowburn}},\ }\bibfield  {title}
	{\bibinfo {title} {{Magneto-optical Kerr effect analysis of magnetic
				nanostructures}},\ }\href {https://doi.org/10.1088/0022-3727/36/18/001}
	{\bibfield  {journal} {\bibinfo  {journal} {J. Phys. D}\ }\textbf {\bibinfo
			{volume} {36}},\ \bibinfo {pages} {2175} (\bibinfo {year}
		{2003})}\BibitemShut {NoStop}%
	\bibitem [{\citenamefont {Kato}\ \emph
		{et~al.}(2004{\natexlab{b}})\citenamefont {Kato}, \citenamefont {Myers},
		\citenamefont {Gossard},\ and\ \citenamefont {Awschalom}}]{Kato2004}%
	\BibitemOpen
	\bibfield  {author} {\bibinfo {author} {\bibfnamefont {Y.~K.}\ \bibnamefont
			{Kato}}, \bibinfo {author} {\bibfnamefont {R.~C.}\ \bibnamefont {Myers}},
		\bibinfo {author} {\bibfnamefont {A.~C.}\ \bibnamefont {Gossard}},\ and\
		\bibinfo {author} {\bibfnamefont {D.~D.}\ \bibnamefont {Awschalom}},\
	}\bibfield  {title} {\bibinfo {title} {{Observation of the Spin Hall Effect
				in Semiconductors}},\ }\href {https://dx.doi.org/10.1126/science.1105514}
	{\bibfield  {journal} {\bibinfo  {journal} {Science}\ }\textbf {\bibinfo
			{volume} {306}},\ \bibinfo {pages} {1910} (\bibinfo {year}
		{2004}{\natexlab{b}})}\BibitemShut {NoStop}%
	\bibitem [{\citenamefont {Zhang}\ \emph {et~al.}(2009)\citenamefont {Zhang},
		\citenamefont {H{\"u}bner}, \citenamefont {Lefkidis}, \citenamefont {Bai},\
		and\ \citenamefont {George}}]{Thomas09np}%
	\BibitemOpen
	\bibfield  {author} {\bibinfo {author} {\bibfnamefont {G.}~\bibnamefont
			{Zhang}}, \bibinfo {author} {\bibfnamefont {W.}~\bibnamefont {H{\"u}bner}},
		\bibinfo {author} {\bibfnamefont {G.}~\bibnamefont {Lefkidis}}, \bibinfo
		{author} {\bibfnamefont {Y.}~\bibnamefont {Bai}},\ and\ \bibinfo {author}
		{\bibfnamefont {T.~F.}\ \bibnamefont {George}},\ }\bibfield  {title}
	{\bibinfo {title} {{Paradigm of the time-resolved magneto-optical Kerr effect
				for femtosecond magnetism}},\ }\href {https://doi.org/10.1038/nphys1315}
	{\bibfield  {journal} {\bibinfo  {journal} {Nat. Phys.}\ }\textbf {\bibinfo
			{volume} {5}},\ \bibinfo {pages} {499} (\bibinfo {year} {2009})}\BibitemShut
	{NoStop}%
	\bibitem [{\citenamefont {Stamm}\ \emph {et~al.}(2017)\citenamefont {Stamm},
		\citenamefont {Murer}, \citenamefont {Berritta}, \citenamefont {Feng},
		\citenamefont {Gabureac}, \citenamefont {Oppeneer},\ and\ \citenamefont
		{Gambardella}}]{Stamm17prl}%
	\BibitemOpen
	\bibfield  {author} {\bibinfo {author} {\bibfnamefont {C.}~\bibnamefont
			{Stamm}}, \bibinfo {author} {\bibfnamefont {C.}~\bibnamefont {Murer}},
		\bibinfo {author} {\bibfnamefont {M.}~\bibnamefont {Berritta}}, \bibinfo
		{author} {\bibfnamefont {J.}~\bibnamefont {Feng}}, \bibinfo {author}
		{\bibfnamefont {M.}~\bibnamefont {Gabureac}}, \bibinfo {author}
		{\bibfnamefont {P.~M.}\ \bibnamefont {Oppeneer}},\ and\ \bibinfo {author}
		{\bibfnamefont {P.}~\bibnamefont {Gambardella}},\ }\bibfield  {title}
	{\bibinfo {title} {{Magneto-Optical Detection of the Spin Hall Effect in Pt
				and W Thin Films}},\ }\href {https://doi.org/10.1103/PhysRevLett.119.087203}
	{\bibfield  {journal} {\bibinfo  {journal} {Phys. Rev. Lett.}\ }\textbf
		{\bibinfo {volume} {119}},\ \bibinfo {pages} {087203} (\bibinfo {year}
		{2017})}\BibitemShut {NoStop}%
	\bibitem [{\citenamefont {Choi}\ \emph {et~al.}(2023)\citenamefont {Choi},
		\citenamefont {Jo}, \citenamefont {Ko}, \citenamefont {Go}, \citenamefont
		{Kim}, \citenamefont {Park}, \citenamefont {Kim}, \citenamefont {Min},
		\citenamefont {Choi},\ and\ \citenamefont {Lee}}]{Choi23nature}%
	\BibitemOpen
	\bibfield  {author} {\bibinfo {author} {\bibfnamefont {Y.-G.}\ \bibnamefont
			{Choi}}, \bibinfo {author} {\bibfnamefont {D.}~\bibnamefont {Jo}}, \bibinfo
		{author} {\bibfnamefont {K.-H.}\ \bibnamefont {Ko}}, \bibinfo {author}
		{\bibfnamefont {D.}~\bibnamefont {Go}}, \bibinfo {author} {\bibfnamefont
			{K.-H.}\ \bibnamefont {Kim}}, \bibinfo {author} {\bibfnamefont {H.~G.}\
			\bibnamefont {Park}}, \bibinfo {author} {\bibfnamefont {C.}~\bibnamefont
			{Kim}}, \bibinfo {author} {\bibfnamefont {B.-C.}\ \bibnamefont {Min}},
		\bibinfo {author} {\bibfnamefont {G.-M.}\ \bibnamefont {Choi}},\ and\
		\bibinfo {author} {\bibfnamefont {H.-W.}\ \bibnamefont {Lee}},\ }\bibfield
	{title} {\bibinfo {title} {{Observation of the orbital Hall effect in a light
				metal Ti}},\ }\href {https://doi.org/10.1038/s41586-023-06101-9} {\bibfield
		{journal} {\bibinfo  {journal} {Nature}\ }\textbf {\bibinfo {volume} {619}},\
		\bibinfo {pages} {52} (\bibinfo {year} {2023})}\BibitemShut {NoStop}%
	\bibitem [{\citenamefont {Lyalin}\ \emph {et~al.}(2023)\citenamefont {Lyalin},
		\citenamefont {Alikhah}, \citenamefont {Berritta}, \citenamefont {Oppeneer},\
		and\ \citenamefont {Kawakami}}]{Lyalin23prl}%
	\BibitemOpen
	\bibfield  {author} {\bibinfo {author} {\bibfnamefont {I.}~\bibnamefont
			{Lyalin}}, \bibinfo {author} {\bibfnamefont {S.}~\bibnamefont {Alikhah}},
		\bibinfo {author} {\bibfnamefont {M.}~\bibnamefont {Berritta}}, \bibinfo
		{author} {\bibfnamefont {P.~M.}\ \bibnamefont {Oppeneer}},\ and\ \bibinfo
		{author} {\bibfnamefont {R.~K.}\ \bibnamefont {Kawakami}},\ }\bibfield
	{title} {\bibinfo {title} {{Magneto-Optical Detection of the Orbital Hall
				Effect in Chromium}},\ }\href
	{https://doi.org/10.1103/PhysRevLett.131.156702} {\bibfield  {journal}
		{\bibinfo  {journal} {Phys. Rev. Lett.}\ }\textbf {\bibinfo {volume} {131}},\
		\bibinfo {pages} {156702} (\bibinfo {year} {2023})}\BibitemShut {NoStop}%
	\bibitem [{Sup()}]{Supp}%
	\BibitemOpen
	\href@noop {} {\bibinfo  {journal} {See Supplemental Material at xx.xxxxxx
			for experimental and theoretical details, which includes Refs.
			\cite{Kao2022,Stamm17prl,Choi23nature,Lyalin23prl,Lyalin24prb,Xie24small,YangYS22,Hurd72book,Ashcroft76book,Haug08boo,YangSY23prl,GaoY14prl,GaoY21prl,YangSY21prl}}\
	}\BibitemShut {NoStop}%
	\bibitem [{\citenamefont {Puri}\ \emph {et~al.}(2024)\citenamefont {Puri},
		\citenamefont {Patel}, \citenamefont {Cabellos}, \citenamefont
		{Rosas-Hernandez}, \citenamefont {Reynolds}, \citenamefont {Churchill},
		\citenamefont {Barraza-Lopez}, \citenamefont {Mendoza},\ and\ \citenamefont
		{Nakamura}}]{Puri24NL}%
	\BibitemOpen
	\bibfield  {journal} {  }\bibfield  {author} {\bibinfo {author} {\bibfnamefont
			{S.}~\bibnamefont {Puri}}, \bibinfo {author} {\bibfnamefont {S.}~\bibnamefont
			{Patel}}, \bibinfo {author} {\bibfnamefont {J.~L.}\ \bibnamefont {Cabellos}},
		\bibinfo {author} {\bibfnamefont {L.~E.}\ \bibnamefont {Rosas-Hernandez}},
		\bibinfo {author} {\bibfnamefont {K.}~\bibnamefont {Reynolds}}, \bibinfo
		{author} {\bibfnamefont {H.~O.}\ \bibnamefont {Churchill}}, \bibinfo {author}
		{\bibfnamefont {S.}~\bibnamefont {Barraza-Lopez}}, \bibinfo {author}
		{\bibfnamefont {B.~S.}\ \bibnamefont {Mendoza}},\ and\ \bibinfo {author}
		{\bibfnamefont {H.}~\bibnamefont {Nakamura}},\ }\bibfield  {title} {\bibinfo
		{title} {{Substrate Interference and Strain in the Second-Harmonic Generation
				from MoSe2 Monolayers}},\ }\href
	{https://doi.org/10.1021/acs.nanolett.4c03880} {\bibfield  {journal}
		{\bibinfo  {journal} {Nano Lett.}\ }\textbf {\bibinfo {volume} {24}},\
		\bibinfo {pages} {13061} (\bibinfo {year} {2024})}\BibitemShut {NoStop}%
	\bibitem [{\citenamefont {Li}\ \emph {et~al.}(2024)\citenamefont {Li},
		\citenamefont {Liu}, \citenamefont {Ye}, \citenamefont {Pan}, \citenamefont
		{Xu}, \citenamefont {Zhu}, \citenamefont {Wang}, \citenamefont {Watanabe},
		\citenamefont {Taniguchi},\ and\ \citenamefont {Liao}}]{LiaoZM24prb2}%
	\BibitemOpen
	\bibfield  {author} {\bibinfo {author} {\bibfnamefont {D.}~\bibnamefont
			{Li}}, \bibinfo {author} {\bibfnamefont {X.-Y.}\ \bibnamefont {Liu}},
		\bibinfo {author} {\bibfnamefont {X.-G.}\ \bibnamefont {Ye}}, \bibinfo
		{author} {\bibfnamefont {Z.-C.}\ \bibnamefont {Pan}}, \bibinfo {author}
		{\bibfnamefont {W.-Z.}\ \bibnamefont {Xu}}, \bibinfo {author} {\bibfnamefont
			{P.-F.}\ \bibnamefont {Zhu}}, \bibinfo {author} {\bibfnamefont {A.-Q.}\
			\bibnamefont {Wang}}, \bibinfo {author} {\bibfnamefont {K.}~\bibnamefont
			{Watanabe}}, \bibinfo {author} {\bibfnamefont {T.}~\bibnamefont
			{Taniguchi}},\ and\ \bibinfo {author} {\bibfnamefont {Z.-M.}\ \bibnamefont
			{Liao}},\ }\bibfield  {title} {\bibinfo {title} {{Facilitating field-free
				perpendicular magnetization switching with a Berry curvature dipole in a Weyl
				semimetal}},\ }\href {https://doi.org/10.1103/PhysRevB.110.L100409}
	{\bibfield  {journal} {\bibinfo  {journal} {Phys. Rev. B}\ }\textbf {\bibinfo
			{volume} {110}},\ \bibinfo {pages} {L100409} (\bibinfo {year}
		{2024})}\BibitemShut {NoStop}%
	\bibitem [{\citenamefont {Li}\ \emph {et~al.}(2025)\citenamefont {Li},
		\citenamefont {Wang}, \citenamefont {Ouyang}, \citenamefont {Li},
		\citenamefont {Yan}, \citenamefont {Dai}, \citenamefont {Shang},
		\citenamefont {Zhang}, \citenamefont {Zhu}, \citenamefont {Li},\ and\
		\citenamefont {Hu}}]{LI25MT}%
	\BibitemOpen
	\bibfield  {author} {\bibinfo {author} {\bibfnamefont {Y.}~\bibnamefont
			{Li}}, \bibinfo {author} {\bibfnamefont {L.}~\bibnamefont {Wang}}, \bibinfo
		{author} {\bibfnamefont {Y.}~\bibnamefont {Ouyang}}, \bibinfo {author}
		{\bibfnamefont {D.}~\bibnamefont {Li}}, \bibinfo {author} {\bibfnamefont
			{Y.}~\bibnamefont {Yan}}, \bibinfo {author} {\bibfnamefont {K.}~\bibnamefont
			{Dai}}, \bibinfo {author} {\bibfnamefont {L.}~\bibnamefont {Shang}}, \bibinfo
		{author} {\bibfnamefont {J.}~\bibnamefont {Zhang}}, \bibinfo {author}
		{\bibfnamefont {L.}~\bibnamefont {Zhu}}, \bibinfo {author} {\bibfnamefont
			{Y.}~\bibnamefont {Li}},\ and\ \bibinfo {author} {\bibfnamefont
			{Z.}~\bibnamefont {Hu}},\ }\bibfield  {title} {\bibinfo {title}
		{{Extraordinary piezoresponse in free-standing two-dimensional Bi2O2Se
				semiconductor toward high-performance light perception synapse}},\ }\href
	{https://doi.org/https://doi.org/10.1016/j.mattod.2024.11.003} {\bibfield
		{journal} {\bibinfo  {journal} {Mater. Today}\ }\textbf {\bibinfo {volume}
			{83}},\ \bibinfo {pages} {12} (\bibinfo {year} {2025})}\BibitemShut {NoStop}%
	\bibitem [{\citenamefont {Zhang}\ \emph {et~al.}(2025)\citenamefont {Zhang},
		\citenamefont {Zobelli}, \citenamefont {Gao}, \citenamefont {Cheng},
		\citenamefont {Zhang}, \citenamefont {Caillaux}, \citenamefont {Qiu},
		\citenamefont {Li}, \citenamefont {Cattelan}, \citenamefont {Kandyba},
		\citenamefont {Barinov}, \citenamefont {Zaghrioui}, \citenamefont
		{Bendounan}, \citenamefont {Rueff}, \citenamefont {Qi}, \citenamefont
		{Perfetti}, \citenamefont {Papalazarou}, \citenamefont {Marsi},\ and\
		\citenamefont {Chen}}]{Zhang25nc}%
	\BibitemOpen
	\bibfield  {author} {\bibinfo {author} {\bibfnamefont {Z.}~\bibnamefont
			{Zhang}}, \bibinfo {author} {\bibfnamefont {A.}~\bibnamefont {Zobelli}},
		\bibinfo {author} {\bibfnamefont {C.}~\bibnamefont {Gao}}, \bibinfo {author}
		{\bibfnamefont {Y.}~\bibnamefont {Cheng}}, \bibinfo {author} {\bibfnamefont
			{J.}~\bibnamefont {Zhang}}, \bibinfo {author} {\bibfnamefont
			{J.}~\bibnamefont {Caillaux}}, \bibinfo {author} {\bibfnamefont
			{L.}~\bibnamefont {Qiu}}, \bibinfo {author} {\bibfnamefont {S.}~\bibnamefont
			{Li}}, \bibinfo {author} {\bibfnamefont {M.}~\bibnamefont {Cattelan}},
		\bibinfo {author} {\bibfnamefont {V.}~\bibnamefont {Kandyba}}, \bibinfo
		{author} {\bibfnamefont {A.}~\bibnamefont {Barinov}}, \bibinfo {author}
		{\bibfnamefont {M.}~\bibnamefont {Zaghrioui}}, \bibinfo {author}
		{\bibfnamefont {A.}~\bibnamefont {Bendounan}}, \bibinfo {author}
		{\bibfnamefont {J.-P.}\ \bibnamefont {Rueff}}, \bibinfo {author}
		{\bibfnamefont {W.}~\bibnamefont {Qi}}, \bibinfo {author} {\bibfnamefont
			{L.}~\bibnamefont {Perfetti}}, \bibinfo {author} {\bibfnamefont
			{E.}~\bibnamefont {Papalazarou}}, \bibinfo {author} {\bibfnamefont
			{M.}~\bibnamefont {Marsi}},\ and\ \bibinfo {author} {\bibfnamefont
			{Z.}~\bibnamefont {Chen}},\ }\bibfield  {title} {\bibinfo {title} {{Rotation
				symmetry mismatch and interlayer hybridization in MoS2-black phosphorus van
				der Waals heterostructures}},\ }\href
	{https://doi.org/10.1038/s41467-025-56113-4} {\bibfield  {journal} {\bibinfo
			{journal} {Nat. Commun.}\ }\textbf {\bibinfo {volume} {16}},\ \bibinfo
		{pages} {763} (\bibinfo {year} {2025})}\BibitemShut {NoStop}%
	\bibitem [{\citenamefont {Xie}\ \emph {et~al.}(2024)\citenamefont {Xie},
		\citenamefont {Zhao}, \citenamefont {Yu},\ and\ \citenamefont
		{Wang}}]{Xie24small}%
	\BibitemOpen
	\bibfield  {author} {\bibinfo {author} {\bibfnamefont {Z.}~\bibnamefont
			{Xie}}, \bibinfo {author} {\bibfnamefont {T.}~\bibnamefont {Zhao}}, \bibinfo
		{author} {\bibfnamefont {X.}~\bibnamefont {Yu}},\ and\ \bibinfo {author}
		{\bibfnamefont {J.}~\bibnamefont {Wang}},\ }\bibfield  {title} {\bibinfo
		{title} {{Nonlinear Optical Properties of 2D Materials and their
				Applications}},\ }\href
	{https://doi.org/https://doi.org/10.1002/smll.202311621} {\bibfield
		{journal} {\bibinfo  {journal} {Small}\ }\textbf {\bibinfo {volume} {20}},\
		\bibinfo {pages} {2311621} (\bibinfo {year} {2024})}\BibitemShut {NoStop}%
	\bibitem [{\citenamefont {Haug}\ and\ \citenamefont
		{Jauho}(2008)}]{Haug08book}%
	\BibitemOpen
	\bibfield  {author} {\bibinfo {author} {\bibfnamefont {H.}~\bibnamefont
			{Haug}}\ and\ \bibinfo {author} {\bibfnamefont {A.-P.}\ \bibnamefont
			{Jauho}},\ }\href@noop {} {\emph {\bibinfo {title} {{Quantum kinetics in
					transport and optics of semiconductors}}}}\ (\bibinfo  {publisher}
	{Springer},\ \bibinfo {year} {2008})\BibitemShut {NoStop}%
	\bibitem [{\citenamefont {Wang}\ \emph
		{et~al.}(2024{\natexlab{b}})\citenamefont {Wang}, \citenamefont {Huang},
		\citenamefont {Liu}, \citenamefont {Feng}, \citenamefont {Zhu}, \citenamefont
		{Wu}, \citenamefont {Xiao},\ and\ \citenamefont {Yang}}]{YangSY24prl}%
	\BibitemOpen
	\bibfield  {author} {\bibinfo {author} {\bibfnamefont {H.}~\bibnamefont
			{Wang}}, \bibinfo {author} {\bibfnamefont {Y.-X.}\ \bibnamefont {Huang}},
		\bibinfo {author} {\bibfnamefont {H.}~\bibnamefont {Liu}}, \bibinfo {author}
		{\bibfnamefont {X.}~\bibnamefont {Feng}}, \bibinfo {author} {\bibfnamefont
			{J.}~\bibnamefont {Zhu}}, \bibinfo {author} {\bibfnamefont {W.}~\bibnamefont
			{Wu}}, \bibinfo {author} {\bibfnamefont {C.}~\bibnamefont {Xiao}},\ and\
		\bibinfo {author} {\bibfnamefont {S.~A.}\ \bibnamefont {Yang}},\ }\bibfield
	{title} {\bibinfo {title} {{Orbital Origin of the Intrinsic Planar Hall
				Effect}},\ }\href {https://doi.org/10.1103/PhysRevLett.132.056301} {\bibfield
		{journal} {\bibinfo  {journal} {Phys. Rev. Lett.}\ }\textbf {\bibinfo
			{volume} {132}},\ \bibinfo {pages} {056301} (\bibinfo {year}
		{2024}{\natexlab{b}})}\BibitemShut {NoStop}%
	\bibitem [{\citenamefont {Misner}\ \emph {et~al.}(1973)\citenamefont {Misner},
		\citenamefont {Thorne},\ and\ \citenamefont {Wheeler}}]{KPThorne73book}%
	\BibitemOpen
	\bibfield  {author} {\bibinfo {author} {\bibfnamefont {C.~W.}\ \bibnamefont
			{Misner}}, \bibinfo {author} {\bibfnamefont {K.~S.}\ \bibnamefont {Thorne}},\
		and\ \bibinfo {author} {\bibfnamefont {J.~A.}\ \bibnamefont {Wheeler}},\
	}\href@noop {} {\emph {\bibinfo {title} {{Gravitation}}}}\ (\bibinfo
	{publisher} {Macmillan},\ \bibinfo {year} {1973})\BibitemShut {NoStop}%
	\bibitem [{\citenamefont {Smith}\ \emph {et~al.}(2022)\citenamefont {Smith},
		\citenamefont {Pullasseri},\ and\ \citenamefont {Srivastava}}]{Smith22prr}%
	\BibitemOpen
	\bibfield  {author} {\bibinfo {author} {\bibfnamefont {T.~B.}\ \bibnamefont
			{Smith}}, \bibinfo {author} {\bibfnamefont {L.}~\bibnamefont {Pullasseri}},\
		and\ \bibinfo {author} {\bibfnamefont {A.}~\bibnamefont {Srivastava}},\
	}\bibfield  {title} {\bibinfo {title} {{Momentum-space gravity from the
				quantum geometry and entropy of Bloch electrons}},\ }\href
	{https://doi.org/10.1103/PhysRevResearch.4.013217} {\bibfield  {journal}
		{\bibinfo  {journal} {Phys. Rev. Res.}\ }\textbf {\bibinfo {volume} {4}},\
		\bibinfo {pages} {013217} (\bibinfo {year} {2022})}\BibitemShut {NoStop}%
	\bibitem [{\citenamefont {Mehraeen}(2025)}]{Mehraeen25prl}%
	\BibitemOpen
	\bibfield  {author} {\bibinfo {author} {\bibfnamefont {M.}~\bibnamefont
			{Mehraeen}},\ }\bibfield  {title} {\bibinfo {title} {{Quantum Response Theory
				and Momentum-Space Gravity}},\ }\href {https://doi.org/10.1103/t6nt-qzws}
	{\bibfield  {journal} {\bibinfo  {journal} {Phys. Rev. Lett.}\ }\textbf
		{\bibinfo {volume} {135}},\ \bibinfo {pages} {156302} (\bibinfo {year}
		{2025})}\BibitemShut {NoStop}%
	\bibitem [{\citenamefont {Onishi}\ \emph {et~al.}(2026)\citenamefont {Onishi},
		\citenamefont {Paul},\ and\ \citenamefont {Fu}}]{FuL26prb}%
	\BibitemOpen
	\bibfield  {author} {\bibinfo {author} {\bibfnamefont {Y.}~\bibnamefont
			{Onishi}}, \bibinfo {author} {\bibfnamefont {N.}~\bibnamefont {Paul}},\ and\
		\bibinfo {author} {\bibfnamefont {L.}~\bibnamefont {Fu}},\ }\bibfield
	{title} {\bibinfo {title} {{Emergent curved space and gravitational lensing
				in quantum materials}},\ }\href {https://doi.org/10.1103/qxnw-8q4y}
	{\bibfield  {journal} {\bibinfo  {journal} {Phys. Rev. B}\ }\textbf {\bibinfo
			{volume} {113}},\ \bibinfo {pages} {024401} (\bibinfo {year}
		{2026})}\BibitemShut {NoStop}%
	\bibitem [{\citenamefont {Soluyanov}\ \emph {et~al.}(2015)\citenamefont
		{Soluyanov}, \citenamefont {Gresch}, \citenamefont {Wang}, \citenamefont
		{Wu}, \citenamefont {Troyer}, \citenamefont {Dai},\ and\ \citenamefont
		{Bernevig}}]{DaiX15nature}%
	\BibitemOpen
	\bibfield  {author} {\bibinfo {author} {\bibfnamefont {A.~A.}\ \bibnamefont
			{Soluyanov}}, \bibinfo {author} {\bibfnamefont {D.}~\bibnamefont {Gresch}},
		\bibinfo {author} {\bibfnamefont {Z.}~\bibnamefont {Wang}}, \bibinfo {author}
		{\bibfnamefont {Q.}~\bibnamefont {Wu}}, \bibinfo {author} {\bibfnamefont
			{M.}~\bibnamefont {Troyer}}, \bibinfo {author} {\bibfnamefont
			{X.}~\bibnamefont {Dai}},\ and\ \bibinfo {author} {\bibfnamefont {B.~A.}\
			\bibnamefont {Bernevig}},\ }\bibfield  {title} {\bibinfo {title} {{Type-II
				Weyl semimetals}},\ }\href {https://dx.doi.org/10.1038/nature15768}
	{\bibfield  {journal} {\bibinfo  {journal} {Nature}\ }\textbf {\bibinfo
			{volume} {527}},\ \bibinfo {pages} {495} (\bibinfo {year}
		{2015})}\BibitemShut {NoStop}%
	\bibitem [{\citenamefont {Wang}\ \emph {et~al.}(2016)\citenamefont {Wang},
		\citenamefont {Liu}, \citenamefont {Liu}, \citenamefont {Pan}, \citenamefont
		{Zhang}, \citenamefont {Zeng}, \citenamefont {Fu}, \citenamefont {Wang},
		\citenamefont {Xu}, \citenamefont {Huang} \emph {et~al.}}]{WanXG16nc}%
	\BibitemOpen
	\bibfield  {author} {\bibinfo {author} {\bibfnamefont {Y.}~\bibnamefont
			{Wang}}, \bibinfo {author} {\bibfnamefont {E.}~\bibnamefont {Liu}}, \bibinfo
		{author} {\bibfnamefont {H.}~\bibnamefont {Liu}}, \bibinfo {author}
		{\bibfnamefont {Y.}~\bibnamefont {Pan}}, \bibinfo {author} {\bibfnamefont
			{L.}~\bibnamefont {Zhang}}, \bibinfo {author} {\bibfnamefont
			{J.}~\bibnamefont {Zeng}}, \bibinfo {author} {\bibfnamefont {Y.}~\bibnamefont
			{Fu}}, \bibinfo {author} {\bibfnamefont {M.}~\bibnamefont {Wang}}, \bibinfo
		{author} {\bibfnamefont {K.}~\bibnamefont {Xu}}, \bibinfo {author}
		{\bibfnamefont {Z.}~\bibnamefont {Huang}}, \emph {et~al.},\ }\bibfield
	{title} {\bibinfo {title} {{Gate-tunable negative longitudinal
				magnetoresistance in the predicted type-II Weyl semimetal
				${\mathrm{WTe}}_{2}$}},\ }\href {https://dx.doi.org/10.1038/ncomms13142}
	{\bibfield  {journal} {\bibinfo  {journal} {Nat. Commun.}\ }\textbf {\bibinfo
			{volume} {7}},\ \bibinfo {pages} {13142} (\bibinfo {year}
		{2016})}\BibitemShut {NoStop}%
	\bibitem [{\citenamefont {Xu}\ \emph {et~al.}(2018)\citenamefont {Xu},
		\citenamefont {Ma}, \citenamefont {Shen}, \citenamefont {Fatemi},
		\citenamefont {Wu}, \citenamefont {Chang}, \citenamefont {Chang},
		\citenamefont {Valdivia}, \citenamefont {Chan}, \citenamefont {Gibson} \emph
		{et~al.}}]{XuSY18np}%
	\BibitemOpen
	\bibfield  {author} {\bibinfo {author} {\bibfnamefont {S.-Y.}\ \bibnamefont
			{Xu}}, \bibinfo {author} {\bibfnamefont {Q.}~\bibnamefont {Ma}}, \bibinfo
		{author} {\bibfnamefont {H.}~\bibnamefont {Shen}}, \bibinfo {author}
		{\bibfnamefont {V.}~\bibnamefont {Fatemi}}, \bibinfo {author} {\bibfnamefont
			{S.}~\bibnamefont {Wu}}, \bibinfo {author} {\bibfnamefont {T.-R.}\
			\bibnamefont {Chang}}, \bibinfo {author} {\bibfnamefont {G.}~\bibnamefont
			{Chang}}, \bibinfo {author} {\bibfnamefont {A.~M.~M.}\ \bibnamefont
			{Valdivia}}, \bibinfo {author} {\bibfnamefont {C.-K.}\ \bibnamefont {Chan}},
		\bibinfo {author} {\bibfnamefont {Q.~D.}\ \bibnamefont {Gibson}}, \emph
		{et~al.},\ }\bibfield  {title} {\bibinfo {title} {{Electrically switchable
				Berry curvature dipole in the monolayer topological insulator
				${\mathrm{WTe}}_{2}$}},\ }\href
	{https://dx.doi.org/10.1038/s41567-018-0189-6} {\bibfield  {journal}
		{\bibinfo  {journal} {Nat. Phys.}\ }\textbf {\bibinfo {volume} {14}},\
		\bibinfo {pages} {900} (\bibinfo {year} {2018})}\BibitemShut {NoStop}%
	\bibitem [{\citenamefont {Qiang}\ \emph {et~al.}(2023)\citenamefont {Qiang},
		\citenamefont {Du}, \citenamefont {Lu},\ and\ \citenamefont {Xie}}]{Lu23prb}%
	\BibitemOpen
	\bibfield  {author} {\bibinfo {author} {\bibfnamefont {X.-B.}\ \bibnamefont
			{Qiang}}, \bibinfo {author} {\bibfnamefont {Z.~Z.}\ \bibnamefont {Du}},
		\bibinfo {author} {\bibfnamefont {H.-Z.}\ \bibnamefont {Lu}},\ and\ \bibinfo
		{author} {\bibfnamefont {X.~C.}\ \bibnamefont {Xie}},\ }\bibfield  {title}
	{\bibinfo {title} {{Topological and disorder corrections to the transverse
				Wiedemann-Franz law and Mott relation in kagome magnets and Dirac
				materials}},\ }\href {https://doi.org/10.1103/PhysRevB.107.L161302}
	{\bibfield  {journal} {\bibinfo  {journal} {Phys. Rev. B}\ }\textbf {\bibinfo
			{volume} {107}},\ \bibinfo {pages} {L161302} (\bibinfo {year}
		{2023})}\BibitemShut {NoStop}%
	\bibitem [{\citenamefont {Nagaosa}\ \emph {et~al.}(2010)\citenamefont
		{Nagaosa}, \citenamefont {Sinova}, \citenamefont {Onoda}, \citenamefont
		{MacDonald},\ and\ \citenamefont {Ong}}]{Nagaosa10rmp}%
	\BibitemOpen
	\bibfield  {author} {\bibinfo {author} {\bibfnamefont {N.}~\bibnamefont
			{Nagaosa}}, \bibinfo {author} {\bibfnamefont {J.}~\bibnamefont {Sinova}},
		\bibinfo {author} {\bibfnamefont {S.}~\bibnamefont {Onoda}}, \bibinfo
		{author} {\bibfnamefont {A.~H.}\ \bibnamefont {MacDonald}},\ and\ \bibinfo
		{author} {\bibfnamefont {N.~P.}\ \bibnamefont {Ong}},\ }\bibfield  {title}
	{\bibinfo {title} {{Anomalous Hall effect}},\ }\href
	{https://doi.org/10.1103/RevModPhys.82.1539} {\bibfield  {journal} {\bibinfo
			{journal} {Rev. Mod. Phys.}\ }\textbf {\bibinfo {volume} {82}},\ \bibinfo
		{pages} {1539} (\bibinfo {year} {2010})}\BibitemShut {NoStop}%
	\bibitem [{\citenamefont {Gong}\ \emph
		{et~al.}(2025{\natexlab{a}})\citenamefont {Gong}, \citenamefont {Du},
		\citenamefont {Sun}, \citenamefont {Lu},\ and\ \citenamefont
		{Xie}}]{GongZH24arXiv}%
	\BibitemOpen
	\bibfield  {author} {\bibinfo {author} {\bibfnamefont {Z.-H.}\ \bibnamefont
			{Gong}}, \bibinfo {author} {\bibfnamefont {Z.~Z.}\ \bibnamefont {Du}},
		\bibinfo {author} {\bibfnamefont {H.-P.}\ \bibnamefont {Sun}}, \bibinfo
		{author} {\bibfnamefont {H.-Z.}\ \bibnamefont {Lu}},\ and\ \bibinfo {author}
		{\bibfnamefont {X.~C.}\ \bibnamefont {Xie}},\ }\href
	{https://arxiv.org/abs/2410.04995} {\bibinfo {title} {{Nonlinear transport
				theory at the order of quantum metric}}} (\bibinfo {year}
	{2025}{\natexlab{a}})\BibitemShut {NoStop}%
	\bibitem [{\citenamefont {Gong}\ \emph
		{et~al.}(2025{\natexlab{b}})\citenamefont {Gong}, \citenamefont {Wei},
		\citenamefont {Lu},\ and\ \citenamefont {Xie}}]{GongZH25arXiv}%
	\BibitemOpen
	\bibfield  {author} {\bibinfo {author} {\bibfnamefont {Z.-H.}\ \bibnamefont
			{Gong}}, \bibinfo {author} {\bibfnamefont {Z.-H.}\ \bibnamefont {Wei}},
		\bibinfo {author} {\bibfnamefont {H.-Z.}\ \bibnamefont {Lu}},\ and\ \bibinfo
		{author} {\bibfnamefont {X.~C.}\ \bibnamefont {Xie}},\ }\href
	{https://arxiv.org/abs/2510.24239} {\bibinfo {title} {{Identifying geometric
				third-order nonlinear transport in disordered materials}}} (\bibinfo {year}
	{2025}{\natexlab{b}})\BibitemShut {NoStop}%
	\bibitem [{\citenamefont {Kao}\ \emph {et~al.}(2022)\citenamefont {Kao},
		\citenamefont {Muzzio}, \citenamefont {Zhang}, \citenamefont {Zhu},
		\citenamefont {Gobbo}, \citenamefont {Yuan}, \citenamefont {Weber},
		\citenamefont {Rao}, \citenamefont {Li}, \citenamefont {Edgar} \emph
		{et~al.}}]{Kao2022}%
	\BibitemOpen
	\bibfield  {author} {\bibinfo {author} {\bibfnamefont {I.~H.}\ \bibnamefont
			{Kao}}, \bibinfo {author} {\bibfnamefont {R.}~\bibnamefont {Muzzio}},
		\bibinfo {author} {\bibfnamefont {H.}~\bibnamefont {Zhang}}, \bibinfo
		{author} {\bibfnamefont {M.}~\bibnamefont {Zhu}}, \bibinfo {author}
		{\bibfnamefont {J.}~\bibnamefont {Gobbo}}, \bibinfo {author} {\bibfnamefont
			{S.}~\bibnamefont {Yuan}}, \bibinfo {author} {\bibfnamefont {D.}~\bibnamefont
			{Weber}}, \bibinfo {author} {\bibfnamefont {R.}~\bibnamefont {Rao}}, \bibinfo
		{author} {\bibfnamefont {J.}~\bibnamefont {Li}}, \bibinfo {author}
		{\bibfnamefont {J.~H.}\ \bibnamefont {Edgar}}, \emph {et~al.},\ }\bibfield
	{title} {\bibinfo {title} {{Deterministic switching of a perpendicularly
				polarized magnet using unconventional spin-orbit torques in
				${\mathrm{WTe}}_{2}$}},\ }\href {https://doi.org/10.1038/s41563-022-01275-5}
	{\bibfield  {journal} {\bibinfo  {journal} {Nat. Mater.}\ }\textbf {\bibinfo
			{volume} {21}},\ \bibinfo {pages} {1029} (\bibinfo {year}
		{2022})}\BibitemShut {NoStop}%
	\bibitem [{\citenamefont {Lyalin}\ and\ \citenamefont
		{Kawakami}(2024)}]{Lyalin24prb}%
	\BibitemOpen
	\bibfield  {author} {\bibinfo {author} {\bibfnamefont {I.}~\bibnamefont
			{Lyalin}}\ and\ \bibinfo {author} {\bibfnamefont {R.~K.}\ \bibnamefont
			{Kawakami}},\ }\bibfield  {title} {\bibinfo {title} {{Interface transparency
				to orbital current}},\ }\href {https://doi.org/10.1103/PhysRevB.110.104418}
	{\bibfield  {journal} {\bibinfo  {journal} {Phys. Rev. B}\ }\textbf {\bibinfo
			{volume} {110}},\ \bibinfo {pages} {104418} (\bibinfo {year}
		{2024})}\BibitemShut {NoStop}%
	\bibitem [{\citenamefont {He}\ \emph {et~al.}(2022)\citenamefont {He},
		\citenamefont {Koon}, \citenamefont {Isobe}, \citenamefont {Tan},
		\citenamefont {Hu}, \citenamefont {Neto}, \citenamefont {Fu},\ and\
		\citenamefont {Yang}}]{YangYS22}%
	\BibitemOpen
	\bibfield  {author} {\bibinfo {author} {\bibfnamefont {P.}~\bibnamefont
			{He}}, \bibinfo {author} {\bibfnamefont {G.}~\bibnamefont {Koon}}, \bibinfo
		{author} {\bibfnamefont {H.}~\bibnamefont {Isobe}}, \bibinfo {author}
		{\bibfnamefont {J.}~\bibnamefont {Tan}}, \bibinfo {author} {\bibfnamefont
			{J.}~\bibnamefont {Hu}}, \bibinfo {author} {\bibfnamefont {A.}~\bibnamefont
			{Neto}}, \bibinfo {author} {\bibfnamefont {L.}~\bibnamefont {Fu}},\ and\
		\bibinfo {author} {\bibfnamefont {H.}~\bibnamefont {Yang}},\ }\bibfield
	{title} {\bibinfo {title} {{Graphene moiré superlattices with giant quantum
				nonlinearity of chiral Bloch electrons}},\ }\href
	{https://doi.org/10.1038/s41565-021-01060-6} {\bibfield  {journal} {\bibinfo
			{journal} {Nat. Nanotechnol.}\ }\textbf {\bibinfo {volume} {17}},\ \bibinfo
		{pages} {378} (\bibinfo {year} {2022})}\BibitemShut {NoStop}%
	\bibitem [{\citenamefont {Hurd}(1972)}]{Hurd72book}%
	\BibitemOpen
	\bibfield  {author} {\bibinfo {author} {\bibnamefont {Hurd}},\ }\href@noop {}
	{\emph {\bibinfo {title} {{The Hall effect in metals and alloys}}}}\
	(\bibinfo  {publisher} {Plenum Press},\ \bibinfo {year} {1972})\BibitemShut
	{NoStop}%
	\bibitem [{\citenamefont {Ashcroft}\ and\ \citenamefont
		{Mermin}(1976)}]{Ashcroft76book}%
	\BibitemOpen
	\bibfield  {author} {\bibinfo {author} {\bibnamefont {Ashcroft}}\ and\
		\bibinfo {author} {\bibnamefont {Mermin}},\ }\href@noop {} {\emph {\bibinfo
			{title} {{Solid state physics}}}}\ (\bibinfo  {publisher} {Thomson
		Learning},\ \bibinfo {year} {1976})\BibitemShut {NoStop}%
	\bibitem [{\citenamefont {Xiao}\ \emph {et~al.}(2023)\citenamefont {Xiao},
		\citenamefont {Wu}, \citenamefont {Wang}, \citenamefont {Huang},
		\citenamefont {Feng}, \citenamefont {Liu}, \citenamefont {Guo}, \citenamefont
		{Niu},\ and\ \citenamefont {Yang}}]{YangSY23prl}%
	\BibitemOpen
	\bibfield  {author} {\bibinfo {author} {\bibfnamefont {C.}~\bibnamefont
			{Xiao}}, \bibinfo {author} {\bibfnamefont {W.}~\bibnamefont {Wu}}, \bibinfo
		{author} {\bibfnamefont {H.}~\bibnamefont {Wang}}, \bibinfo {author}
		{\bibfnamefont {Y.-X.}\ \bibnamefont {Huang}}, \bibinfo {author}
		{\bibfnamefont {X.}~\bibnamefont {Feng}}, \bibinfo {author} {\bibfnamefont
			{H.}~\bibnamefont {Liu}}, \bibinfo {author} {\bibfnamefont {G.-Y.}\
			\bibnamefont {Guo}}, \bibinfo {author} {\bibfnamefont {Q.}~\bibnamefont
			{Niu}},\ and\ \bibinfo {author} {\bibfnamefont {S.~A.}\ \bibnamefont
			{Yang}},\ }\bibfield  {title} {\bibinfo {title} {{Time-Reversal-Even
				Nonlinear Current Induced Spin Polarization}},\ }\href
	{https://doi.org/10.1103/PhysRevLett.130.166302} {\bibfield  {journal}
		{\bibinfo  {journal} {Phys. Rev. Lett.}\ }\textbf {\bibinfo {volume} {130}},\
		\bibinfo {pages} {166302} (\bibinfo {year} {2023})}\BibitemShut {NoStop}%
\end{thebibliography}
\end{document}